\documentclass{llncs}
\usepackage{enumitem}
\usepackage{graphicx}
\usepackage{subcaption}
\usepackage{multirow}
\usepackage{tabularx}
\usepackage{afterpage}
\usepackage{hyperref}

\begin{document}

\title{Reinventing NetFlow for OpenFlow Software-Defined Networks (Technical report)}
\titlerunning{NetFlow/IPFIX in OpenFlow}  
%
\author{Jos\'{e} Su\'{a}rez-Varela\inst{1} \and Pere Barlet-Ros\inst{1,	2}}
\authorrunning{Jos\'{e} Su\'{a}rez-Varela et al.} 
%
\tocauthor{Jos\'{e} Su\'{a}rez-Varela, Pere Barlet-Ros}
\institute{UPC BarcelonaTech, Spain,\\
\email{\{jsuarezv,pbarlet\}@ac.upc.edu}
\and
Talaia Networks, S.L., Spain}

\maketitle              

\begin{abstract}
Obtaining flow-level measurements, similar to those provided by Netflow/IPFIX, with OpenFlow is challenging as it requires the installation of an entry per flow in the flow tables. This approach does not scale well with the number of concurrent flows in the traffic as the number of entries in the flow tables is limited and small. Flow monitoring rules may also interfere with forwarding or other rules already present in the switches, which are often defined at different granularities than the flow level. In this paper, we present a transparent and scalable flow-based monitoring solution that is fully compatible with current off-the-shelf OpenFlow switches. As in NetFlow/IPFIX, we aggregate packets into flows directly in the switches and asynchronously send traffic reports to an external collector. In order to reduce the overhead, we implement three different traffic sampling methods depending on the OpenFlow features available in the switch. We developed our complete flow monitoring solution within OpenDaylight and evaluated its accuracy in a testbed with Open vSwitch. Our experimental results using real-world traffic traces show that the proposed sampling methods are accurate and can effectively reduce the resource requirements of flow measurements in OpenFlow.

\end{abstract}
\section{Introduction and related work}
The paradigm of Software-Defined networking (SDN) has recently gained lots of attention from research and industry. The logically centralized control plane provides flexibility and enables to perform a fine-grained management of the network, taking advantage of the decision making from a global perspective of the network. To be successful in dynamic environments, monitoring takes a key role in SDN given that management applications often need to make use of accurate and timely traffic measurements at different aggregation levels. Specifically, there are many applications such as traffic engineering, anomaly detection, traffic classification, traffic shaping or Service Level Agreement (SLA) enforcing based on the collection of an accurate set of per-flow measurements.

Since its inception in 2008, OpenFlow \cite{openflow} has become a dominant protocol for the southbound interface (between control and data planes) in SDN. It is impossible to foresee whether OpenFlow will ever evolve towards a measurement standard technology, but potentially it could be a definitive solution for traffic measurement. It can maintain records with flow statistics and includes an interface that allows to retrieve measurements at different aggregation levels passively (when a flow entry expires), or actively (by querying the statistics to the switch).

However, an inherent issue of SDN is its scalability. For a proper design of a monitoring system, it is necessary to consider the network and processing overhead to store and gather the flow statistics. On the one hand, it should not be ignored the fact that controllers are critical points in the infrastructure since all the management decisions are made and communicated from there to each switch under its control. On the other hand, the most straightforward way of implementing per-flow monitoring is by maintaining an entry for each flow in a table of the switch. Each of these entries has some counters which are updated every time a packet matches them. Thus, obtaining fine-grained measurements of all flows results in a great constraint, since nowadays OpenFlow commodity switches do not support a large number of flow entries due to their limited hardware resources available (i.e., the number of TCAM entries and processing power) \cite{iStamp}. For the sake of scalability, a common practice in traditional networks is to implement traffic sampling when collecting flow measurements. As for the sampling schemes, two different approaches can be mainly distinguished: packet sampling and flow sampling. The former consists of sampling each packet with a specific probability and aggregating the statistics in different records for each flow\footnote{Interpreting a flow as a set of packets sharing the same IP 5-tuple \{src\_IP, dst\_IP, src\_port, dst\_port, protocol\}}. While the latter consists of sampling a flow with some probability and aggregating all the packets of this flow in a separated record. Packet sampling has been extensively used traditionally, e.g., in NetFlow \cite{netflow}, JFlow \cite{jflow} or sFlow \cite{sflow}. It provides a coarse view of traffic, which is sufficient for applications such as traffic volume estimation or \textit{heavy hitters} detection. However, with this method small flows are underrepresented, if noticed at all. In this paper, we implement flow sampling instead because it is easier to provide without requiring modifications to the OpenFlow specification. Several studies have shown that packet sampling it is not the most adequate solution for some fine-grained monitoring applications \cite{sekar:reiter}. This is particularly the case of applications like traffic classification or anomaly detection, where flow sampling is a better alternative. 

Another approach to address the scalability issues, is to design distributed solutions \cite{opennetmon,dream}. This type of solutions takes advantage of the global view of the network in the controller to compute the active paths. For example, in OpenNetMon \cite{opennetmon}, they design an scheme to monitor flows in edge switches and make measurements of throughput, packet loss and delay. In such way, they overcome the limitation of the small number of TCAM entries available in the switch. However, they still may have severe problems with the overhead in the controller, which has to calculate all the paths and install as equitable as possible the flow entries in all the edge switches in the network.

In the light of the above, we present a monitoring solution which emulates the NetFlow/IPFIX operation with OpenFlow and implements flow sampling. In this way, for each flow sampled, we maintain a flow entry in the switch. Here each entry records the duration (in seconds and nanoseconds) and packet and bytes counts. We use timeouts to define when these records are going to expire and, therefore, being reported to the controller. A similar approach was previously used in \cite{tma} to assess the accuracy of measurements and timeouts in some OpenFlow switches. However, their approach is not scalable as it requires to install an entry in the flow tables for every single flow observed in the traffic, assumes that all rules have been deployed proactively for every flow that will be observed in the network, and does not address the problem of how monitoring rules interfere with the rest of rules installed in the switch (e.g., forwarding rules). In contrast, we present a complete flow monitoring solution that has the following novel features:

\begin{itemize}

\item \textbf{Scalable:} In order to tackle the scalability issues mentioned earlier, our system performs traffic sampling. Particularly, three different sampling methods were designed depending on the OpenFlow features available in the switch. This results in less overhead for the controller and a more reduced number of entries in flow tables than in the case of monitoring all the flows. We remark that our methods only require to initially install some rules in the switch which will operate autonomously to discriminate \(pseudo\) randomly the traffic to be sampled. For each flow sampled, a flow entry is added reactively in the switch to record the per-flow statistics. To the best of our knowledge, there are no solutions in line with this approach. For example, iSTAMP \cite{iStamp} performs a flow-based sampling technique where they make use of a multi-armed-bandit algorithm to ``stamp" the most informative flows and maintain particular entries to record per-flow metrics. Likewise, this solution needs to perform periodically a training phase with some iterations to detect those flows. This means that, for each training phase, it does not work well until the algorithm achieves a proper set of flows. Additionally, this solution specifically addresses the detection of particular flows like \textit{heavy hitters} or specific flow sub-populations, while our solution provides a generic dataset of the flows in the network.

\item \textbf{Fully compliant with OpenFlow:} Our system makes use exclusively of messages and features described in the OpenFlow specification. Specifically, we consider a solution fully compatible with OpenFlow 1.1.0 and later versions. Although we provide a less transparent solution for OpenFlow 1.0.0 since we are aware that it has a large support in current commodity switches. It makes our solution more pragmatic and realistic for current SDN deployments, which strongly relies on OpenFlow. Alternatively, some authors suggest to make use of different architectures or protocols specifically designed for monitoring tasks. For example, in \cite{opensketch}, they propose an architecture called OpenSketch where some sketches can be defined and dynamically loaded to perform different measurement tasks. Likewise, OpenSample \cite{opensample} leverages sFlow \cite{sflow} to perform packet sampling. But, in favor of OpenFlow, it is important to remark that it is a vendor-independent technology with a strong support in the SDN industry. This makes it highly prone to be adopted by all vendors and enable the interoperability among switches. In \cite{opennetmon}, the authors highlight the importance of making an OpenFlow compatible monitoring solution, since it is cheaper to implement and does not require standardization by a larger community. Our solution also has support for IPv6 traffic for switches with OpenFlow 1.2.0 and later.

\item \textbf{Transparent:} Our system can be interpreted as an additional module which does not affect the correct operation of other modules performing different network functions (e.g., traffic forwarding). To ensure it, we make use of the pipeline processing feature with multiple tables of OpenFlow. In such way, our monitoring solution operates in the first table and forwards all the packets to the next table, where other module(s) can install their own flow entries.

\item \textbf{Asynchronous collection of flow statistics:} Our system performs a passive measuring and retrieves flow statistics when the flow expires (either by an idle or hard timeout). In FlowSense \cite{flowsense}, they propose this mechanism to retrieve statistics for all the entries in the switches to estimate per-flow link utilization. The main problem of this solution is that for flows with large timeouts, statistics are retrieved after too long a time. This makes obtaining accurate measurements unfeasible in dynamic environments with highly fluctuating traffic. In our solution, as our module is completely decoupled from other rules with other purposes, we can define adaptively the timeouts to obtain accurate measurements. Although the algorithm to adapt timeouts is out of the scope of this paper, we cite some solutions that could be adopted, like those proposed in PayLess \cite{payless} or OpenNetMon \cite{opennetmon}, where they design adaptive schedule algorithms to perform queries in switches.
\end{itemize}

The remainder of this paper is structured as follows: Firstly, in Section 2, we provide an OpenFlow overview focusing on the features and messages involved in our solution. Section 3 defines our proposed monitoring system. In Section 4, we evaluate our monitoring system in a testbed with Open vSwitch \cite{openvswitch} and an implementation within OpenDaylight \cite{opendaylight}. Here we include an analysis of the accuracy of the three different flow sampling methods proposed and an evaluation of the overhead contribution, both with real-world traffic traces. Lately, in Section 5 we conclude and mention some aspects for future works.

\section{OpenFlow background}

Nowadays, there is a growing trend among vendors to adopt OpenFlow for their switches in two different ways. Some of them are opting for OpenFlow-only devices, while others offer hybrid switches, where both traditional network protocols and OpenFlow coexist. At the moment, the latest version is OpenFlow 1.5.1 (published in April 2015), but it is quite unusual to find commodity switches with higher support than OpenFlow 1.3.0. 

In this section, we particularly focus on the OpenFlow 1.1.0 specification, since it is the first version fully compatible with our solution. This is because from this version it is possible to make use of multiple tables, which enable us to design a transparent system. However, we propose an alternative solution with some limitations for switches with OpenFlow 1.0.0 support (more details will be explained in Section \ref{subsecsec:modularization}). It is also worth mentioning that everything described for our solution can be applied to IPv6 traffic from OpenFlow 1.2.0 onwards. In this case, in line with the OpenFlow specification, all the entries containing match fields for IP protocol or higher layer protocols have to be installed separately for IPv4 and IPv6 as it is mandatory to specify the ethernet type field in the entry.  

Regarding the monitoring solution proposed in this paper, we provide below a summary of the principal elements and messages involved here.

\subsection{Multiple flow tables and groups}

Multiple flow tables and groups are both available from OpenFlow 1.1.0. The support of multiple tables enables to decouple the ruleset of different modules operating in different tables. It introduces a flexible pipeline processing of the packets and it is much more efficient when there are network modules which deploy orthogonal processing of packets (e.g., ACL, QoS and routing), since it avoids to create a large ruleset due to cross product of all the rules.

Packets begin their processing pipeline in the first table of the device and can be directed to other tables. In this way, as it goes through the pipeline, a packet can both execute an action and continue the processing in the next table or accumulate the actions and apply them at the end of the pipeline. In order to resolve possible conflicts between overlapping rules in the same flow table, each entry has a priority field.

Groups are abstractions which allow to represent a set of actions for all packets matching an entry in a flow table. Each group table contains a number of buckets, which in turn are composed by a set of actions. Therefore, if a bucket is selected, all its actions will be applied to the packet. There are four different mechanisms to select the buckets applied to a packet reaching the group table: 1)All (e.g., for multicast), 2)Select (e.g., for multipath), 3)Indirect and 4)Fast Failover (e.g., to use first live port). Our solution leverages the \textit{select} mechanism for the hash-based method described in Section 3.1. In a group of type \textit{select}, packets are processed by a single bucket and so, only actions within the selected bucket are applied. This bucket selection depends on a selection algorithm (external to the OpenFlow specification) implemented in the switch which should perform equal or weighted load sharing among buckets.

\subsection{Adding new flow entries and groups}

When a packet matches an entry in a flow table with an action \textit{output to controller}, a portion of this packet is encapsulated in an OFPT\_PACKET\_IN message and forwarded to the controller. Also, packets are usually sent to the controller when they do not match any rule in the flow table, since switches typically have a default (wilcarded) rule to perform this action. The OFPT\_PACKET\_IN message includes an identification field of the table where the action \textit{output to controller} was executed. It is an important information for our solution since it enables us to filter packets from the table where the monitoring module is operating and treat them in a particular way. Once the packet has been processed, the controller may send an OFPT\_FLOW\_MOD message of type OFPFC\_ADD to the switch to install a new flow entry with a set of instructions. In this way, these instructions will be applied for the subsequent packets matching the particular fields defined in this entry. That is the natural mechanism in OpenFlow networks to add reactively new flows appearing in the switch. In the OFPT\_FLOW\_MOD message, it is possible to specify two timeouts (idle and hard) for that particular entry to define when it is going to be removed from the switch. The idle timeout defines the maximum time interval between two consecutive packets matching this entry, while the hard timeout is the maximum lifetime since the entry was installed.

In order to add a new group, the controller may send an OFPT\_GROUP\_MOD message of type OFPGC\_ADD to the switch. This message defines the type of group (All, Select, Indirect or Fast Failover), a set of buckets with their correspondent actions set and an unique identifier (32 bits) for this group. We remark that a group table does not contain match fields, but only actions within buckets which may be applied for packets directed to this group. In order to forward packets to a group table, it is necessary to add an entry in a flow table (with match fields) defining an action of type OFPAT\_GROUP. This action must include the unique identifier of the group. Likewise, from a group table it is possible to forward packets to another group. 

\subsection{Statistics collection}

To collect flow measurements, two different approaches can be mainly remarked. On the one hand, pull-based mechanisms consist of making active measurements, i.e., sending queries (OFPT\_MULTIPART\_REQUEST message) to the switch for the desired flows. The switch will respond with an OFPT\_MULTIPART\_REPLY message with a summary of the flow (duration in seconds and nanoseconds, packet count and bytes count). This approach is illustrated in OpenNetMon \cite{opennetmon}, where they perform an adaptive polling to collect the data from edge switches. On the other hand, push-based mechanisms consist of collecting measurements asynchronously. In this case, when adding a new flow entry, idle and/or hard timeouts are defined. Then, when a flow entry is evicted, the switch sends to the controller an OFPT\_FLOW\_REMOVED message with the flow statistics. This message also informs with flags that indicate if the expiration was caused by either the idle or the hard timeout. This method is that proposed in FlowSense \cite{flowsense} as a solution for passive measuring with OpenFlow. To receive asynchronously this message, when adding a new flow, the controller has to explicitly note it in the OFPT\_FLOW\_MOD message by marking the flag OFPFF\_SEND\_FLOW\_REM.

\section{Monitoring system}

In this section, we present our monitoring solution, which is implemented within the OpenDaylight controller. Our system fully relies on the OpenFlow specification to emulate the operation of NetFlow/IPFIX in traditional networks. This is not new in SDN, since some works, such as \cite{tma}, used a similar approach earlier. However, to the best of our knowledge, no previous works proposed OpenFlow compatible methods to implement traffic sampling in a NetFlow/IPFIX style, i.e., randomly sampling the traffic and maintaining per-flow statistics in separated records, which are finally reported to a collector. Since we are aware that OpenFlow has many features that are classified as ``\textit{optional}" in the specification, we designed different sampling methods which have a different level of requirements of features available in the switch. These methods, in summary, consist of installing a set of entries in the switch which allow us to discriminate directly the traffic to be sampled. Thus, we only send the first packets of those flows to be monitored and the controller is in charge of installing reactively specific flow entries to sample these flows. Since OpenFlow switches are capable of communicating to the controller the features available, it is possible to decide the method to be used separately for each switch depending on its capabilities. We did not design any method for packet sampling since we found it excessively complex to implement with the current OpenFlow support.

\begin{figure}[!t]
\centering
\begin{subfigure}{.5\textwidth}
  \centering
  \includegraphics[width=0.97\linewidth]{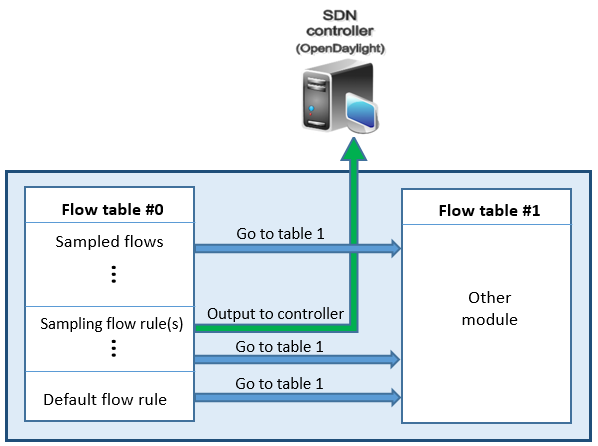}
  \caption{Sampling based on IP suffixes or ports}
  \label{fig:ip-port}
\end{subfigure}%
\begin{subfigure}{.5\textwidth}
  \centering
  \includegraphics[width=0.97\linewidth]{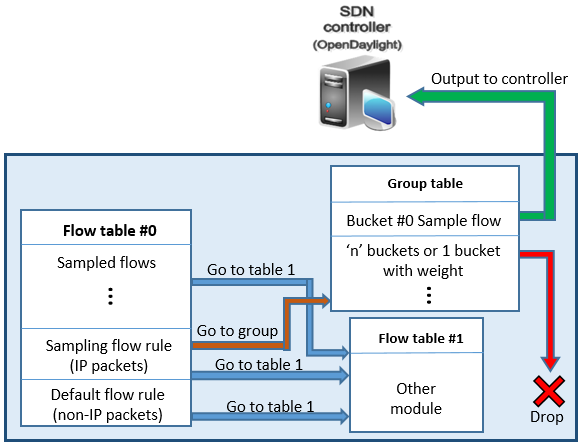}
  \caption{Sampling based on hash function}
  \label{fig:hash}
\end{subfigure}
\caption{Scheme of OpenFlow tables and entries of the monitoring system}
\label{fig:sampling}
\end{figure}

Before showing the details of each method, we describe the generic structure of OpenFlow tables and entries in our system, which is illustrated in Fig.\ref{fig:ip-port}. In all the methods proposed, the monitoring system operates in the first table of the switch (table \#0), where the pipeline process for incoming packets starts. In this way, our system installs in this table some entries to sample the traffic and maintains records for monitored flows. All the entries in this table have at least one instruction to direct the packets to another table, where other entries can be installed for different purposes (e.g., forwarding). Thus, we ensure that all the packets reach the next table and it enables us to completely decouple our monitoring module from others operating in other tables. Focusing in the first table, where our system operates, three different blocks of entries can be differentiated by their priority field. There is a first block of flow level (5-tuple) entries that act as flow records. Then, a block of entries with lower priority which define the packets to be sampled. And lastly, we add a default entry with the lowest priority which simply directs to the next table the packets (including non-IP packets) that did not match any previous entries. In this way, the key point of our system resides on the second block of entries, where the methods described below establish different rules to decide which packets are sampled. The operation mode when a new packet arrives to the switch is to check firstly if it is already in one of the per-flow monitoring entries. If it matches any of these entries, the counters of packets and bytes are updated and the packet is directed to the next table. If not, it goes through the set of entries that define whether it has to be sampled or not. If it matches one of these, then the packet is directed to the next table and also forwarded to the controller to be processed and adding an specific entry in the first block to sample subsequent packets of this flow. Finally, if the packet does not match any of the previous rules, it is directed to the next table.

\subsection{Proposed sampling methods}

We present here the three different sampling methods devised for our monitoring solution and discuss the OpenFlow features required for each of them. We assume that the switches have support for OpenFlow 1.1.0 and later versions so, they have at least support for multiple tables. However, in Section \ref{subsecsec:modularization}, we make some comments about how to implement an alternative solution with OpenFlow 1.0.0. Our selection mechanisms for the packets are covered by the Packet Sampling (PSAMP) Protocol Specification \cite{psamp}, which is compatible with the IPFIX protocol specification to export packet information. According to the PSAMP specification terminology, our first couple of sampling methods can be classified as property match filtering, where a packet is selected if specific fields within the packet are equal to a predefined set of values. While the latest method is of type hash-based filtering, which consists of computing a hash function on some bits of the packet header and to select it if the hash value falls in the hash selection range.

\subsubsection{A. Sampling based on IP suffixes}
This method is based on performing flow sampling based on IP addresses matches. To achieve it, the controller adds proactively one entry in the block of sampling rules with match fields for particular IP addresses ranges. Typically, in traditional routers the matching of IP addresses is based in IP prefixes. In contrast, we consider to apply a mask which checks the last \textit{n} bits of the IPs, i.e., we sample flows with specific IP suffixes. In this way, we sample a more representative set of flows, since we monitor flows from different subnets (IP prefixes) in the network. In order to implement this, it is only necessary a wildcarded entry that filters the IP suffixes desired for source or destination addresses, or combinations of them. To control the number of flows to be sampled we make a rough consideration that, in average, flows are homogeneously distributed along the whole the IP range (we later analyze this assumption with real traffic in Section \ref{subsec:accuracy}). As a consequence, for each bit checked in the mask, the number of flows sampled will be divided by two with respect to the total number of flows arriving to the switch. We are aware that typically there are some IPs that generate much more traffic than others, but this method somehow allow to control the number of flows to be monitored. Furthermore, if we consider pairs of IPs for the selection, instead of individual IPs, we should control better this effect. In this case, if we are sampling hosts which generate a large number of flows, only those flows which match both, source and destination IP suffixes, are sampled. Generically, our sampling rate can be defined by the expression (\ref{eq:ip-suffixes}).
\begin{equation}
sampling\:rate =  \frac{1}{2^m\cdot 2^n}
\label{eq:ip-suffixes}
\end{equation}

Where 'm' is the number of bits checked for the source IP suffix and 'n' the number of bits checked for the destination IP suffix. 

This method is similar to host-based (or host-pair-based) sampling, as we are using IP addresses to select the packets to be sampled. However, host-based schemes typically provide statistics of aggregated traffic for individual or group of hosts. In contrast, we sample the traffic by single or pairs of IP suffixes, but provide individual statistics at a flow granularity level. Moreover, to avoid bias in the selection, the IP suffixes can be periodically changed by simply replacing the sampling rule(s) in the OpenFlow table.

For this method the only \textit{optional} requirement of OpenFlow is the support of arbitrary masks for IP match fields to check suffixes, since there are some switches which only support to make use of prefix masks for IPs.

\subsubsection{B. Sampling based on ports}

This method performs flow sampling based on port matches. We can add proactively entries with match fields of ports for TCP, UDP and SCTP, since OpenFlow has support for these protocols. We make use of entries that check source or destination ports, or combinations of them. In that way, to perform flow sampling as random as possible, we choose randomly \textit{n} ports out of 65,535, which is the total number of possible ports (port fields have 16 bits). Thus, we can control the sampling rate by adding a particular number of ports to be sampled. To achieve it, we can make the rough assumption that, in average, the traffic is homogeneously distributed among all the ports (we later analyze this assumption with real traffic in Section \ref{subsec:accuracy}). As in the previous method, if we consider pairs of ports, this assumption should be more correct. If we want to sample by pairs of ports, two tables can be used to check first the source port and, in case of matching some of them, direct the packet to other table where the destination port is checked. In that way, we avoid to add a large number of entries due to cross-product of all the source and destination port numbers. We remark also that for each protocol (e.g., TCP) it is necessary to add separated entries, since OpenFlow does not have support to add generic match fields for ports without specifying a particular IP protocol type. Generically, we can estimate the sampling rate for pairs of ports with the next formula (\ref{eq:pair-ports}).
\begin{equation}
sampling\:rate =  \frac{m\cdot n}{65,535^2}
\label{eq:pair-ports}
\end{equation}

Where 'm' is the number of entries with different source port match fields and 'm' the number of entries for different destination ports.

To perform a sampling considering only the source or destination port, we can apply the formula (\ref{eq:single-port}).
\begin{equation}
sampling\:rate =  \frac{m}{65,535}
\label{eq:single-port}
\end{equation}

Where 'm' is the number of entries with different source or destination ports match fields.

For this method there are not required \textit{optional} features of OpenFlow. However, we should consider that the number of entries with different port numbers can be very large if we want to sample a great fraction of the flows. For example, if we want to sample approximately 1 out of 200 flows only by source or destination ports, there will be necessary 65,535/200 $\approx$ 328 entries for each protocol to be monitored. For pairs of source and destination ports the number of entries would be increased to 9,268 entries approximately.

\subsubsection{C. Hash-based flow sampling}

This method consists of computing a hash function on the traditional 5-tuple fields of the packet header and selecting it if the hash value falls in a particular range. In Fig.\ref{fig:hash}, we can see the OpenFlow tables structure of this method. In this case, all IP packets are directed to a group table where only one bucket sends the packet to the controller to monitor the flow, other buckets drop the packet. To control the sampling rate, we can select a weight for each bucket. This method much better controls the sampling rate, since we can assume that a hash function is homogeneous along its range for all the different flows in the switch.

The requirements for this method are to support group tables (optionally available from OpenFlow 1.1.0) with \textit{select} buckets and having an accurate algorithm in the switch, which is external to the OpenFlow specification, to balance the load properly among buckets.

\subsection{Modularization of the system}\label{subsecsec:modularization}

As mentioned at the beginning of Section 3, our solution leverages the support of multiple tables to make independent its operation. Thus, we can see our monitoring system as an independent module in the controller which is responsible for the monitoring tasks and does not interfere which other modules operating in other tables. In the controller we can filter and process the Packet In messages triggered by entries of our module, since these messages contain the table Id of the entry which forwarded the packet to the controller. Nevertheless, we propose an alternative for those switches with OpenFlow 1.0.0 support, where only a table can be used. Since this version does not support group tables, only the two first methods, based on matches for IP suffixes and ports, can be implemented. Thus, it is feasible to install the monitoring entries by combining them with the correspondent actions of other modules at the expense of loosing the decoupling of our monitoring system. Furthermore, if we maintain entries at the 5-tuple level to record statistics, it would not be possible to make use of rules finer than this level of aggregation for other purposes, which can be a limitation for the correct operation of other modules.

\subsection{Statistics retrieval}

Our system envisions a push-based approach to retrieve statistics. Given that it uses specific entries, we can selectively choose the timeouts to retrieve the statistics. As a result, we overcome the issue of other push-based solutions such as FlowSense \cite{flowsense}, where flows with large timeouts are collected after too long a time decreasing the accuracy of the measurements. Additionally, we consider the possibility of combining this reactive collection mechanism with a timely polling for those flows with highly fluctuating traffic. So, we can obtain more accurate measurements and avoid forcing the expiration of these monitoring entries very frequently.

\section{Experimental evaluation}

In this section we evaluate our monitoring solution. We have implemented it within OpenDaylight \cite{opendaylight}, operating jointly with the ``L2Switch" \cite{l2switch} module that it includes for layer 2 forwarding.

We conducted experiments in a small testbed with an Open vSwitch \cite{openvswitch}, a host (VM) which injects traffic into the switch and another host which acts as a sink for all the traffic forwarded. All the experiments make use of real traffic traces from a large Spanish university (labeled as ``UNIVERSITY"), MAWI \cite{mawi} and CAIDA \cite{caida}. These traces were filtered to keep only the TCP and UDP traffic. In Table \ref{table:traces} there is a detailed description of each trace.

\setlength{\tabcolsep}{0.1cm}
\renewcommand{\arraystretch}{1.5}

\begin{table}[!ht]
\centering
\resizebox{\textwidth}{!}{%
\begin{tabular}{|c|c|c|c|}
\hline
\textbf{Trace dataset} & \begin{tabular}[c]{@{}c@{}}\textbf{\# of flows}\end{tabular} & \begin{tabular}[c]{@{}c@{}}\textbf{\# of packets}\end{tabular} & \begin{tabular}[c]{@{}c@{}}\textbf{Description}\end{tabular} \\ \hline \hline

\begin{tabular}[c]{@{}c@{}}UNIVERSITY\\ \scriptsize{25th November 2016}\end{tabular} & \begin{tabular}[c]{@{}c@{}}2,972,880 (total flows)\\2,349,677 (TCP flows)\\623,203 (UDP flows)\end{tabular} & 75,585,871 & \begin{tabular}{>{\centering\arraybackslash}p{9cm}} 10 Gbps downstream access link of a large Spanish university, which connects about 25 faculties and 40 departments (geographically distributed in 10 campuses) to the Internet through the Spanish Research and Education network (RedIRIS).\\Average traffic rate: 2.41 Gbps\end{tabular} \\ \hline

\begin{tabular}[c]{@{}c@{}}MAWI\\ \scriptsize{15th July 2016}\end{tabular} & \begin{tabular}[c]{@{}c@{}}3,299,166 (total flows)\\2,653,150 (TCP flows)\\646,016 (UDP flows)\end{tabular} & 54,270,059 & \begin{tabular}{>{\centering\arraybackslash}p{9cm}}1 Gbps transit link of WIDE network to the upstream ISP. Trace from the samplepoint-F.\\Average traffic rate: 507 Mbps\end{tabular}  \\ \hline

\begin{tabular}[c]{@{}c@{}}CAIDA\\ \scriptsize{18th February 2016}\end{tabular} & \begin{tabular}[c]{@{}c@{}}2,353,413 (total flows)\\1,992,983 (TCP flows)\\360,430 (UDP flows)\end{tabular} & 51,368,574 & \begin{tabular}{>{\centering\arraybackslash}p{9cm}}This trace corresponds to a 10 Gbps link of a data center in Chicago (direction A - from Seattle to Chicago).\\Average traffic rate: 2.9 Gbps\end{tabular} \\ \hline

\end{tabular}%
}
\caption{Summary of the traces used in our experiments}
\label{table:traces}
\vspace{-1cm}
\end{table}

\subsection{Accuracy of our flow sampling methods}\label{subsec:accuracy}

We conduct experiments to assess if the sampling rate is applied properly and if the selection of flows is random enough when using the proposed sampling methods. All our experiments were separately done for the MAWI, CAIDA and UNIVERSITY traces described above and repeated applying sampling rates of 1/64, 1/128, 1/256, 1/512 and 1/1024. For the methods based on IP suffixes or ports, we consider two modalities: matching only on a source IP suffix (or a set of source ports), and matching both, source and destination IP suffixes (or ports). For each of these modalities, with a particular trace, and a specific sampling rate, we performed 500 experiments selecting randomly IP suffixes (or ports). For the hash-based method, since it is based on a deterministic selection function, we only conducted one experiment for each case.

\begin{figure}[!ht]
\centering
\begin{subfigure}{0.33\textwidth}
  \centering
  \includegraphics[width=\linewidth, height=2.8cm]{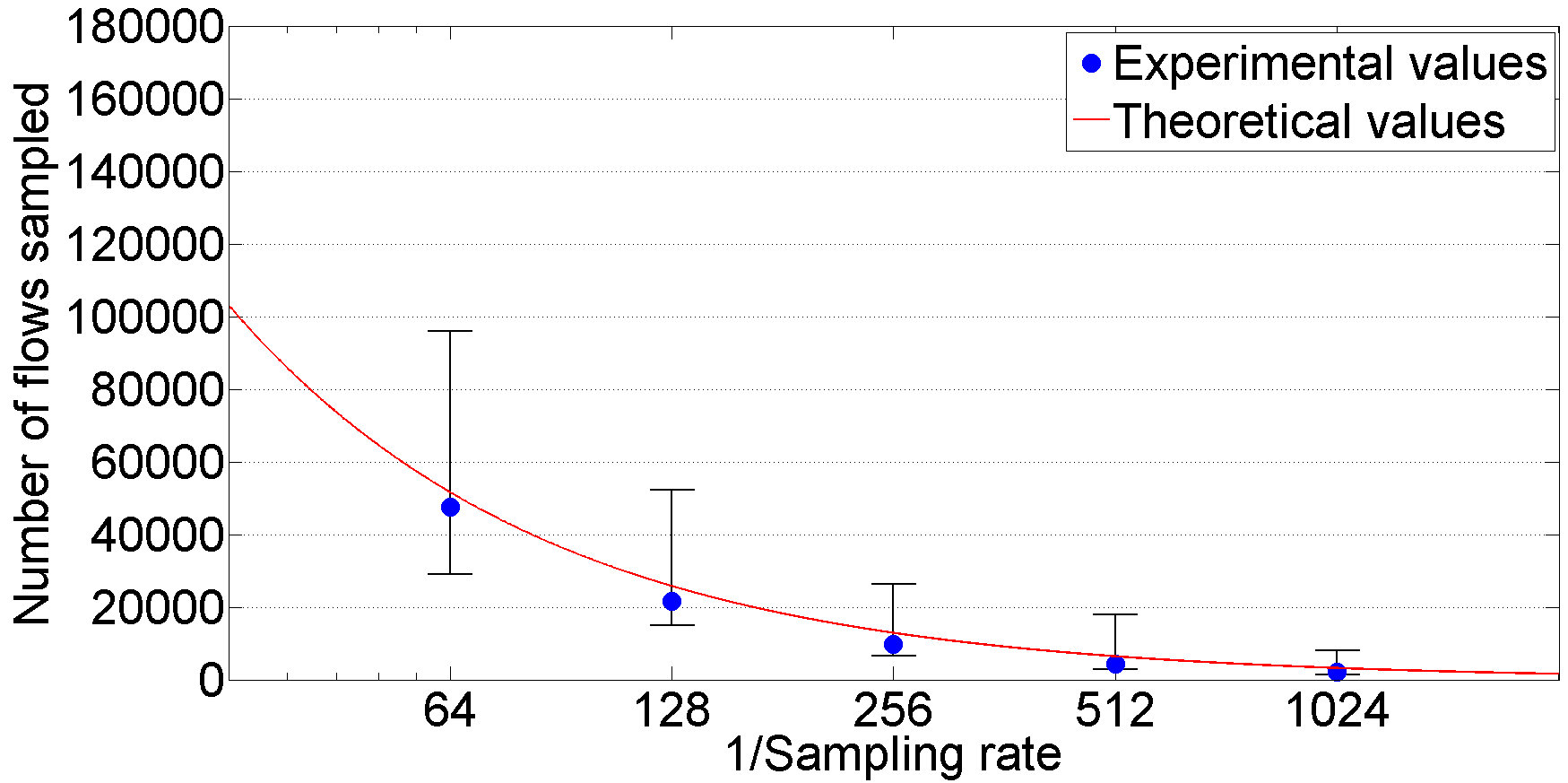}
  \caption{MAWI trace}
  \label{fig:sample-rate-IP-source:MAWI}
\end{subfigure}%
\begin{subfigure}{0.33\textwidth}
  \centering
  \includegraphics[width=\linewidth, height=2.8cm]{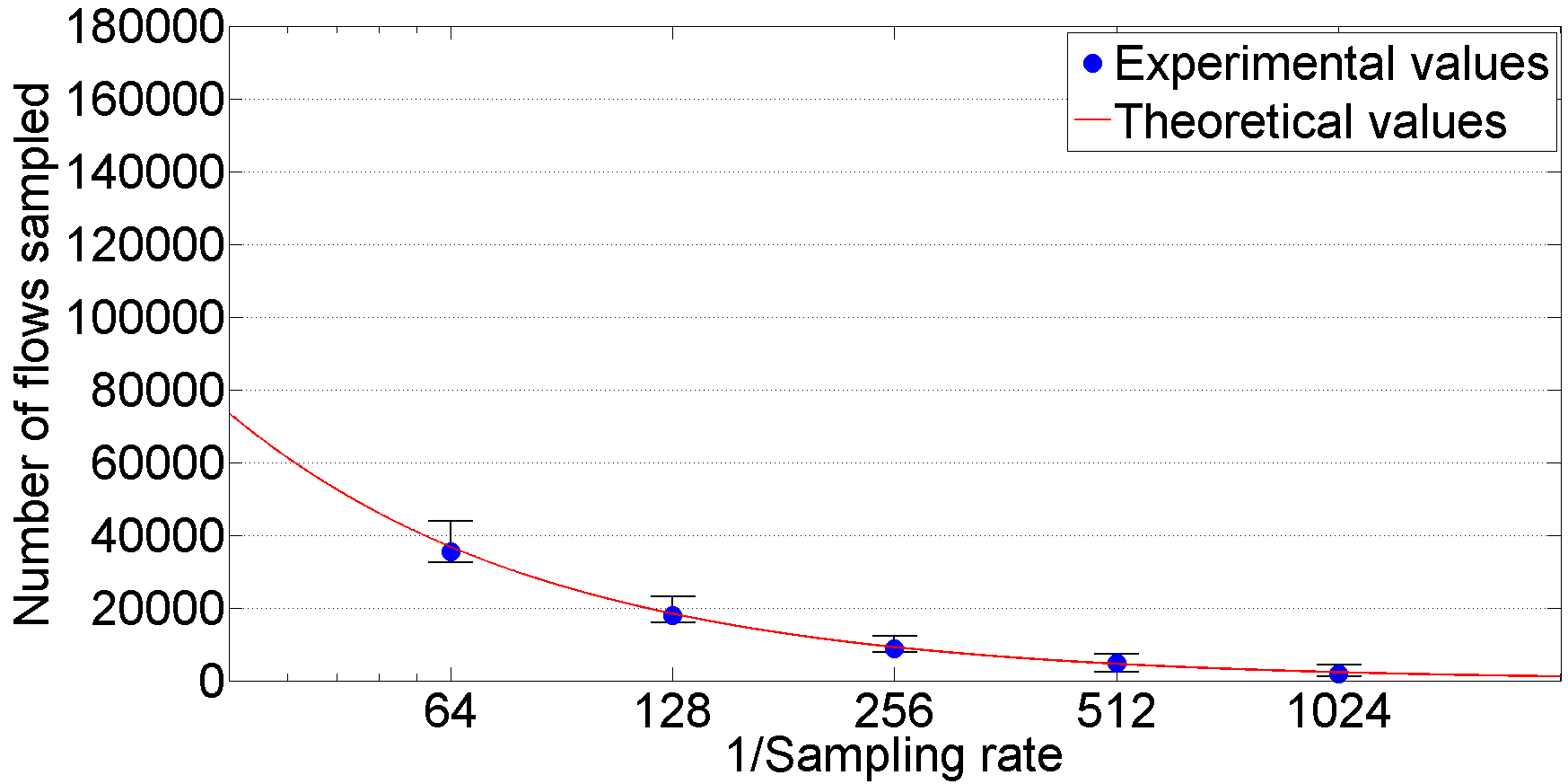}
  \caption{CAIDA trace}
  \label{fig:sample-rate-IP-source:CAIDA}
\end{subfigure}
\begin{subfigure}{0.33\textwidth}
  \centering
  \includegraphics[width=\linewidth, height=2.8cm]{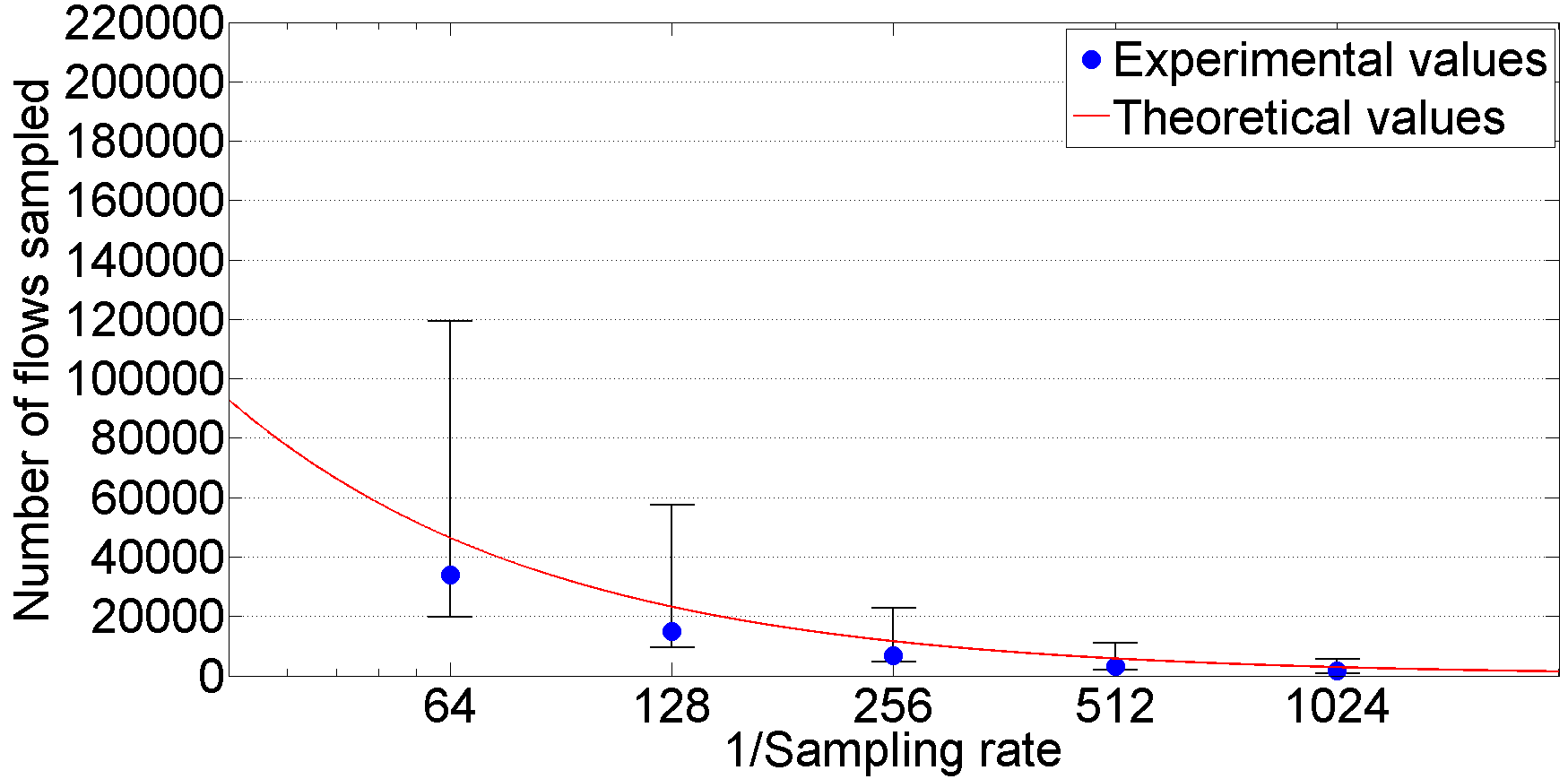}
  \caption{UNIVERSITY trace}
  \label{fig:sample-rate-IP-source:UPC}
\end{subfigure}
\caption{Evaluation of sampling rate for method based on source IP suffixes}
\label{fig:sample-rate-IP-source}
\end{figure}

\begin{figure}[!ht]
\centering
\begin{subfigure}{0.33\textwidth}
  \centering
  \includegraphics[width=\linewidth, height=2.8cm]{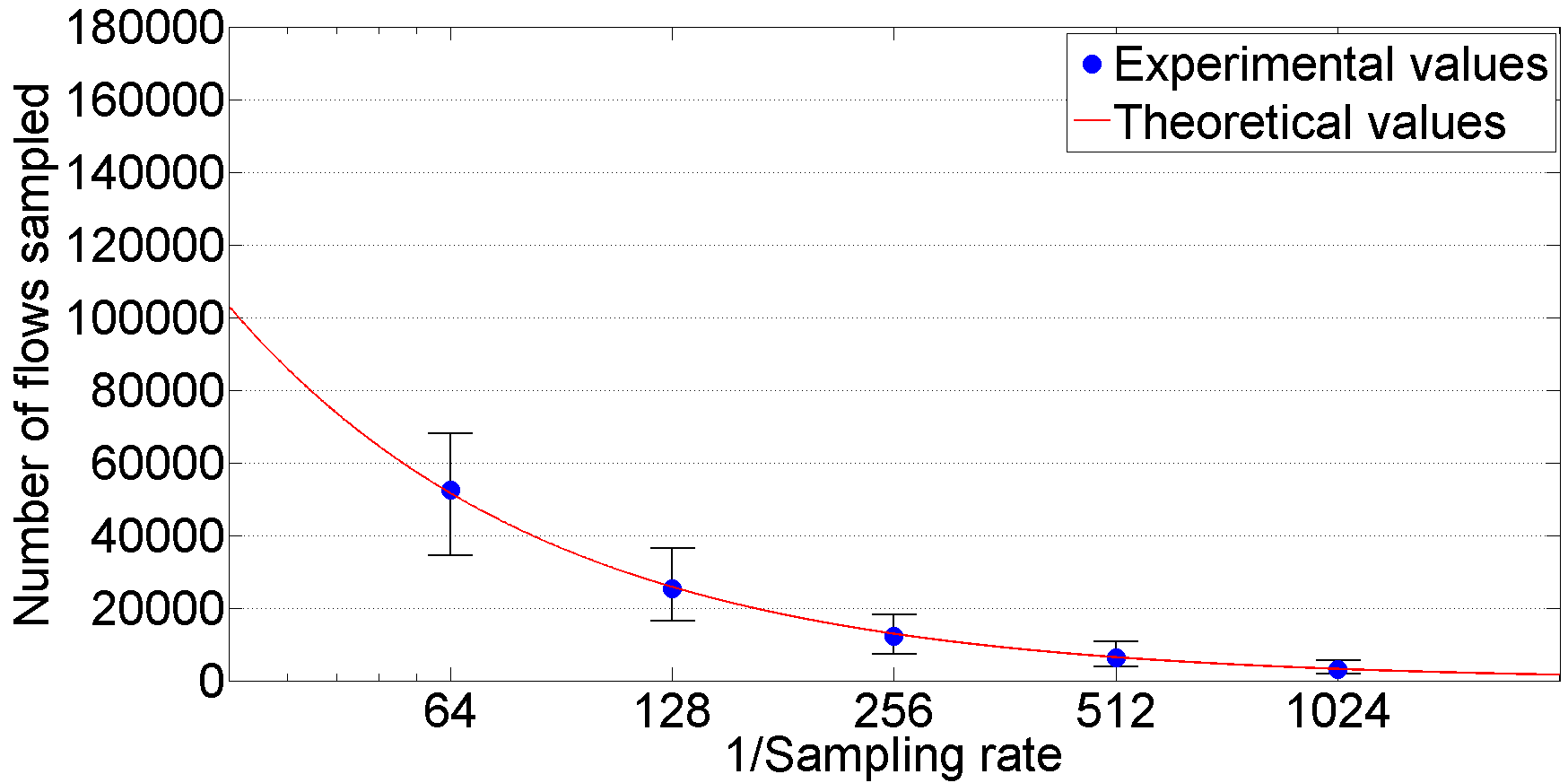}
  \caption{MAWI trace}
  \label{fig:sample-rate-IP-pairs:MAWI}
\end{subfigure}%
\begin{subfigure}{0.33\textwidth}
  \centering
  \includegraphics[width=\linewidth, height=2.8cm]{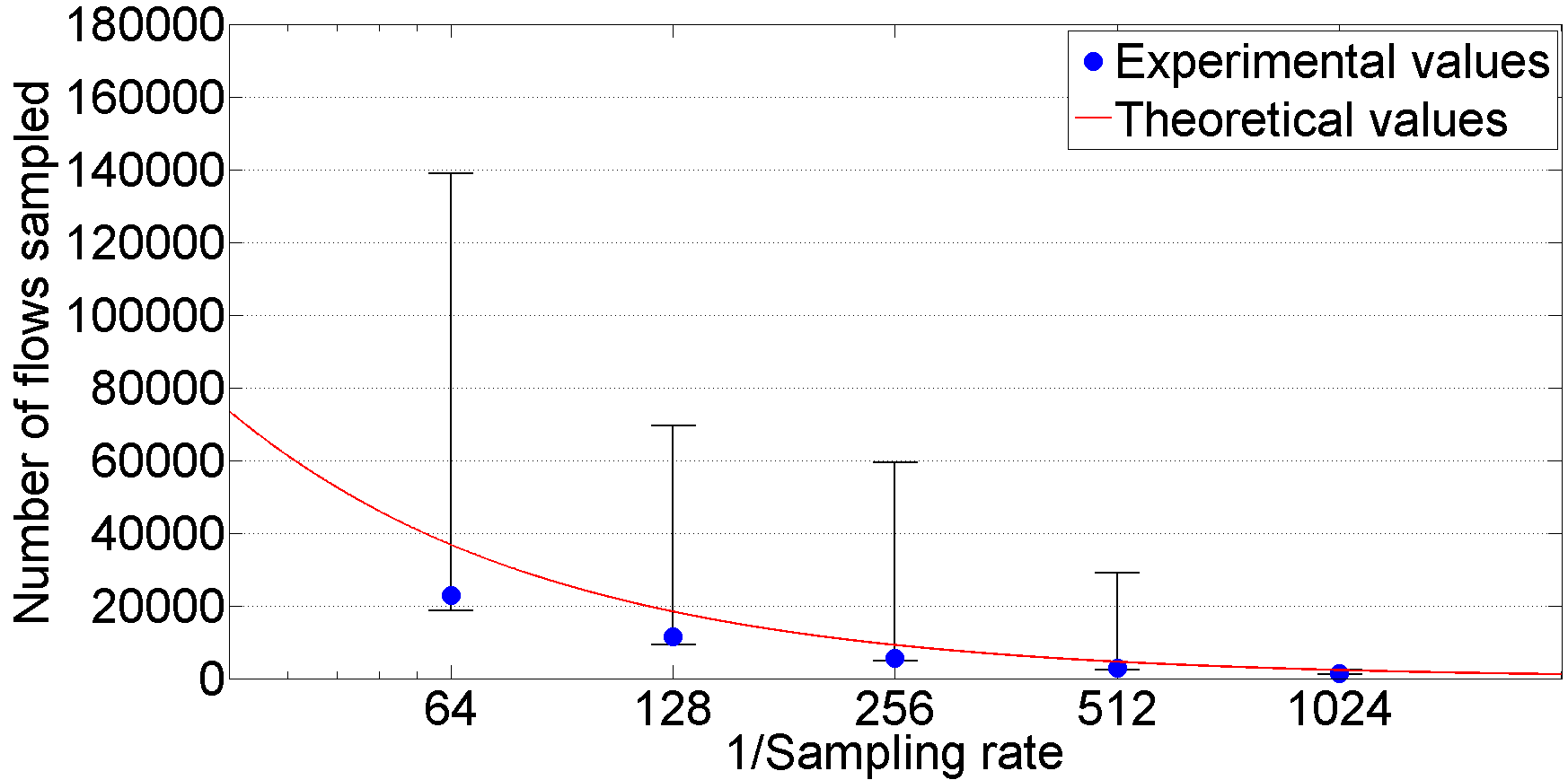}
  \caption{CAIDA trace}
  \label{fig:sample-rate-IP-pairs:CAIDA}
\end{subfigure}
\begin{subfigure}{0.33\textwidth}
  \centering
  \includegraphics[width=\linewidth, height=2.8cm]{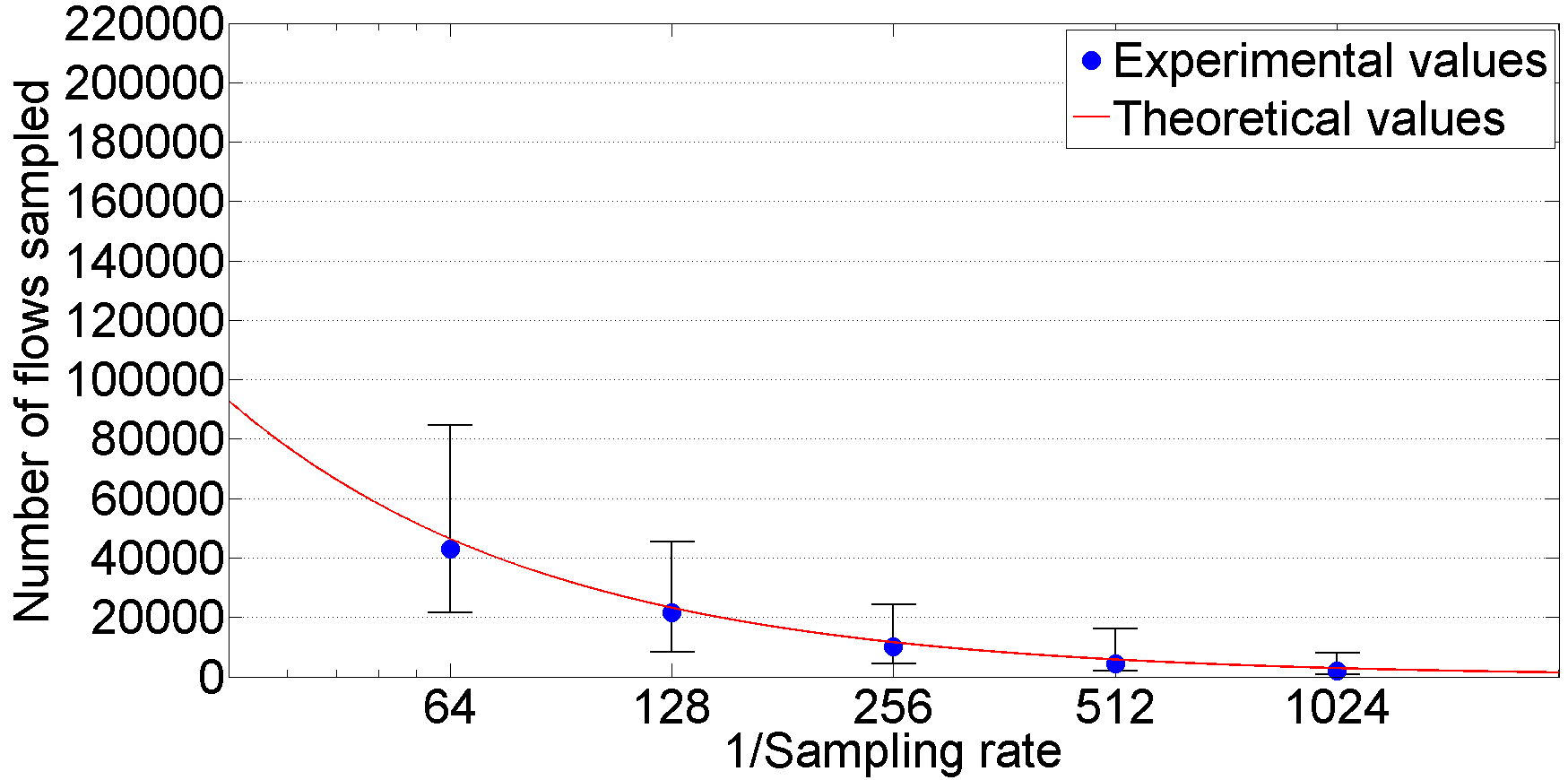}
  \caption{UNIVERSITY trace}
  \label{fig:sample-rate-IP-pairs:UPC}
\end{subfigure}
\caption{Evaluation of sampling rate for method based on pairs of IP suffixes}
\label{fig:sample-rate-pairs}
\end{figure}

\begin{figure}[!ht]
\centering
\begin{subfigure}{0.33\textwidth}
  \centering
  \includegraphics[width=\linewidth, height=2.8cm]{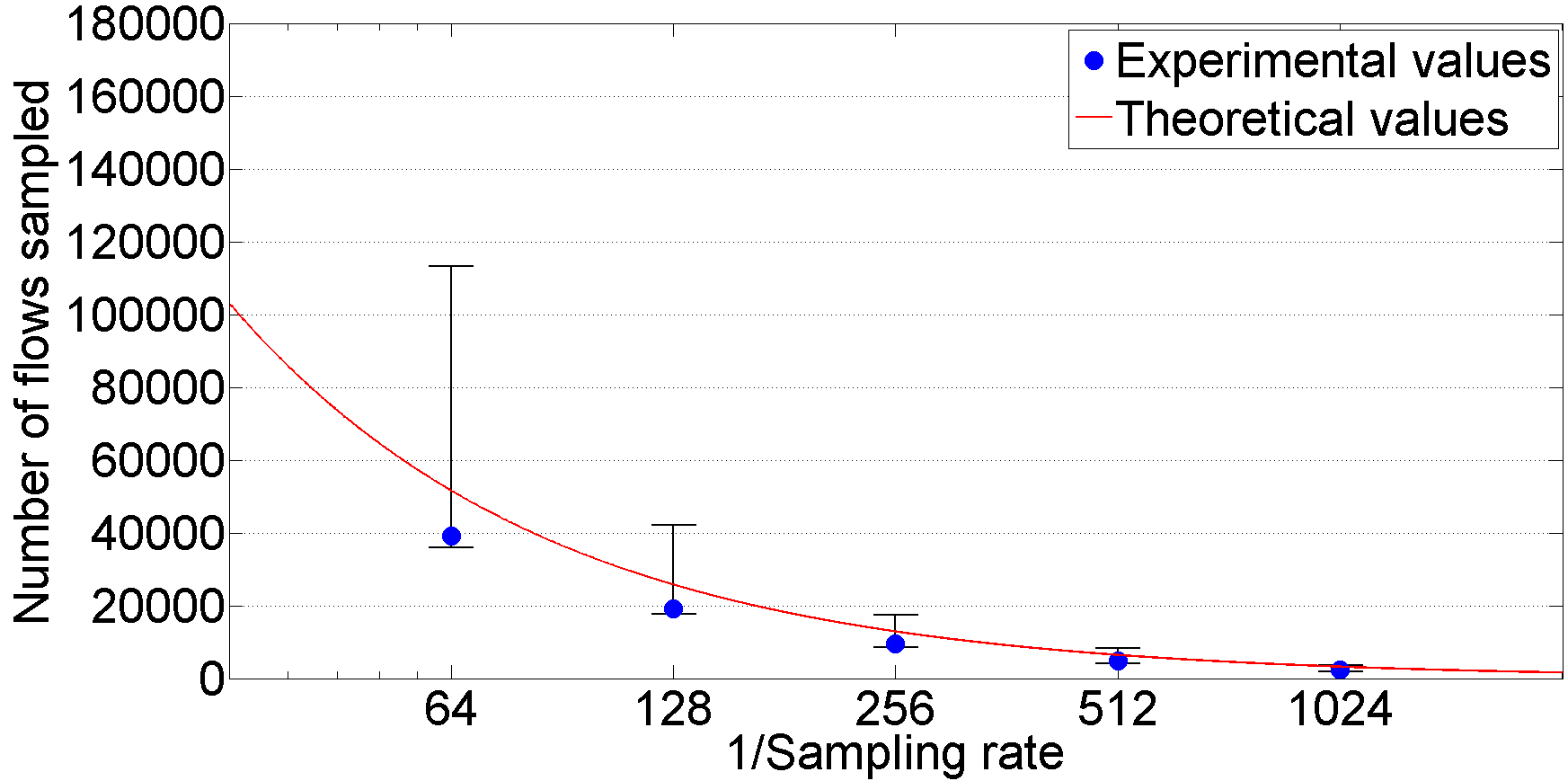}
  \caption{MAWI trace}
  \label{fig:sample-rate-port-source:MAWI}
\end{subfigure}%
\begin{subfigure}{0.33\textwidth}
  \centering
  \includegraphics[width=\linewidth, height=2.8cm]{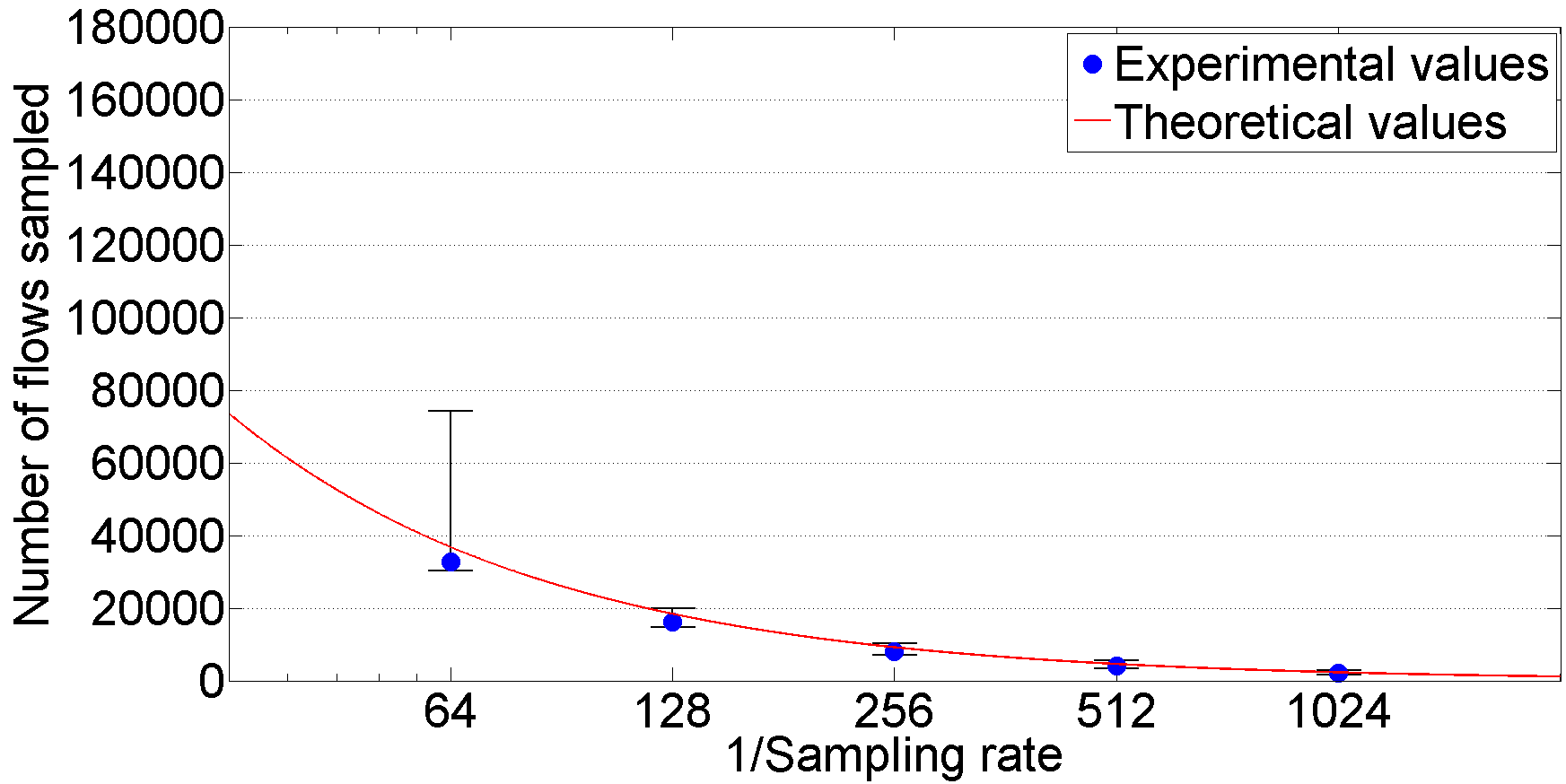}
  \caption{CAIDA trace}
  \label{fig:sample-rate-port-source:CAIDA}
\end{subfigure}
\begin{subfigure}{0.33\textwidth}
  \centering
  \includegraphics[width=\linewidth, height=2.8cm]{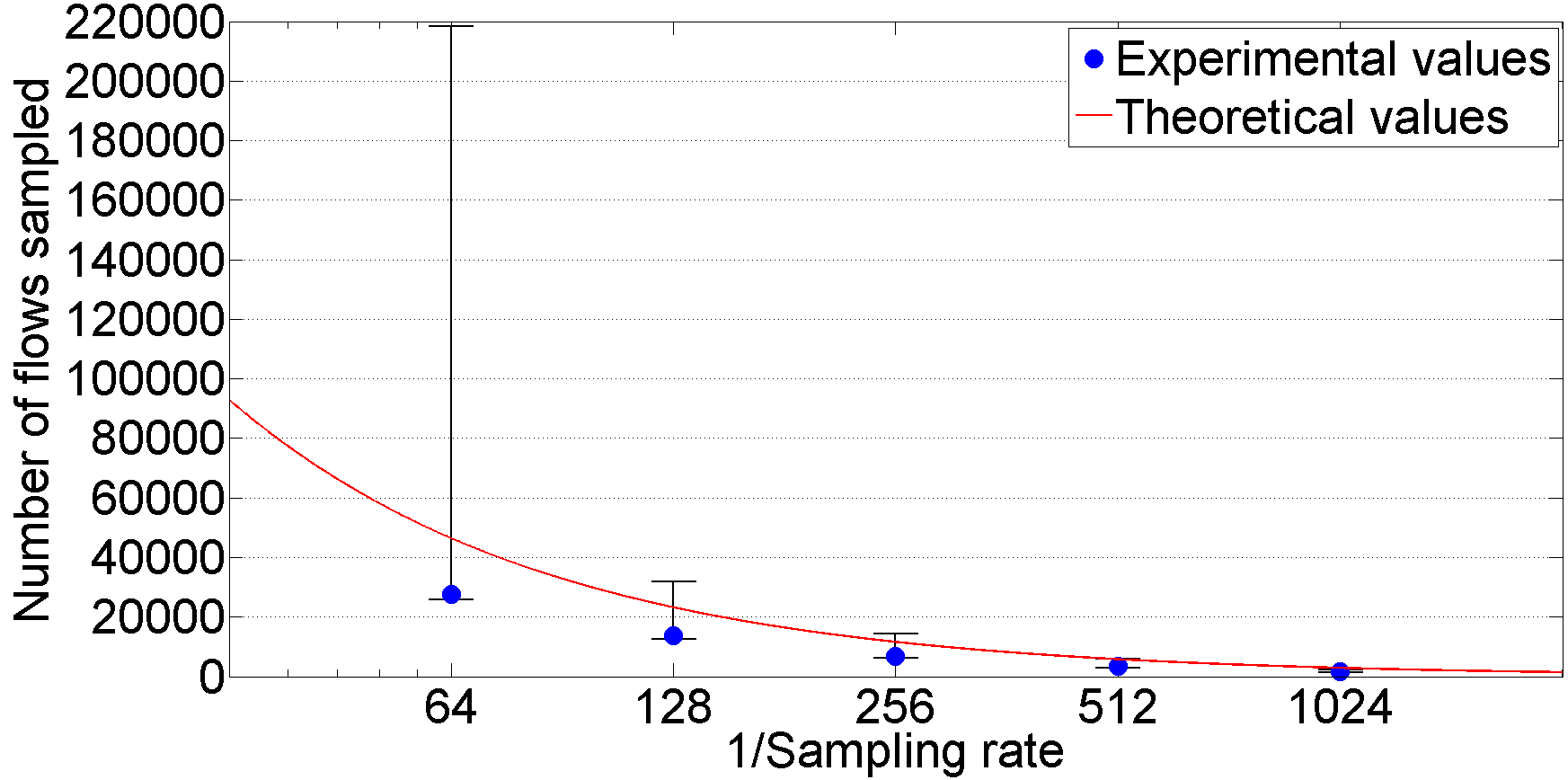}
  \caption{UNIVERSITY trace}
  \label{fig:sample-rate-port-source:UPC}
\end{subfigure}
\caption{Evaluation of sampling rate for method based on source ports}
\label{fig:sample-rate-port-source}
\end{figure}

\begin{figure}[!ht]
\centering
\begin{subfigure}{0.33\textwidth}
  \centering
  \includegraphics[width=\linewidth, height=2.8cm]{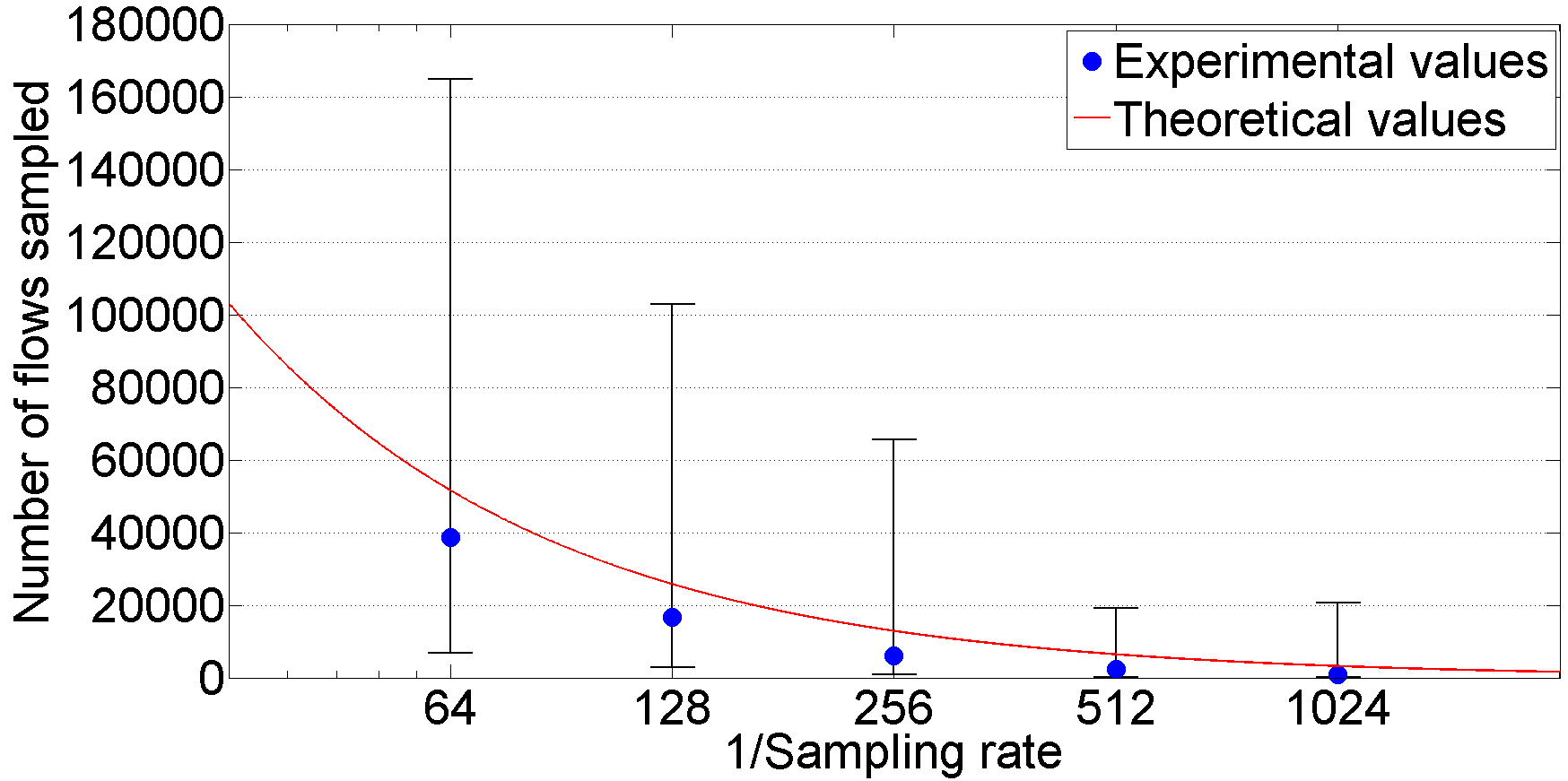}
  \caption{MAWI trace}
  \label{fig:sample-rate-port-pairs:MAWI}
\end{subfigure}%
\begin{subfigure}{0.33\textwidth}
  \centering
  \includegraphics[width=\linewidth, height=2.8cm]{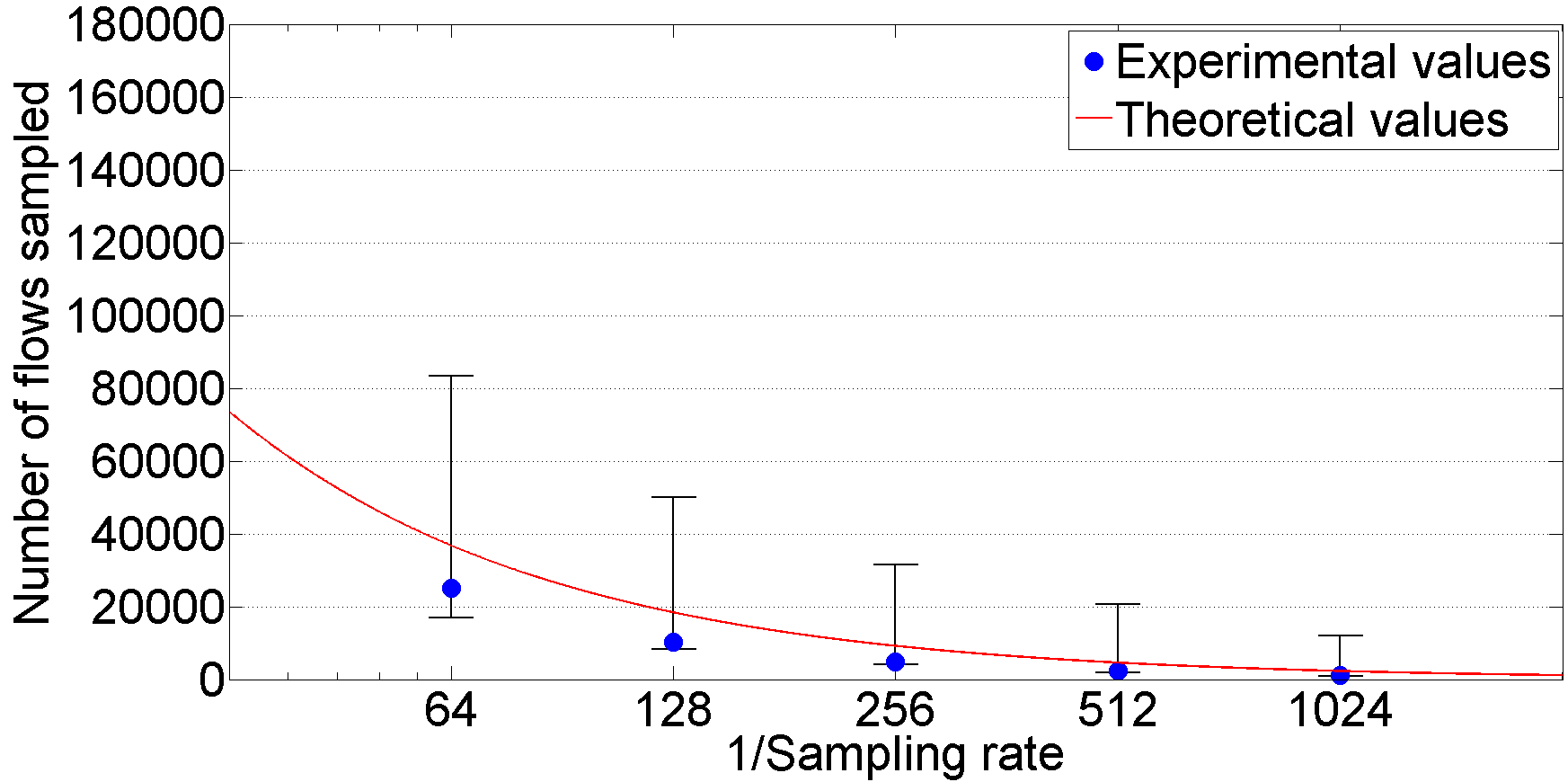}
  \caption{CAIDA trace}
  \label{fig:sample-rate-port-pairs:CAIDA}
\end{subfigure}
\begin{subfigure}{0.33\textwidth}
  \centering
  \includegraphics[width=\linewidth, height=2.8cm]{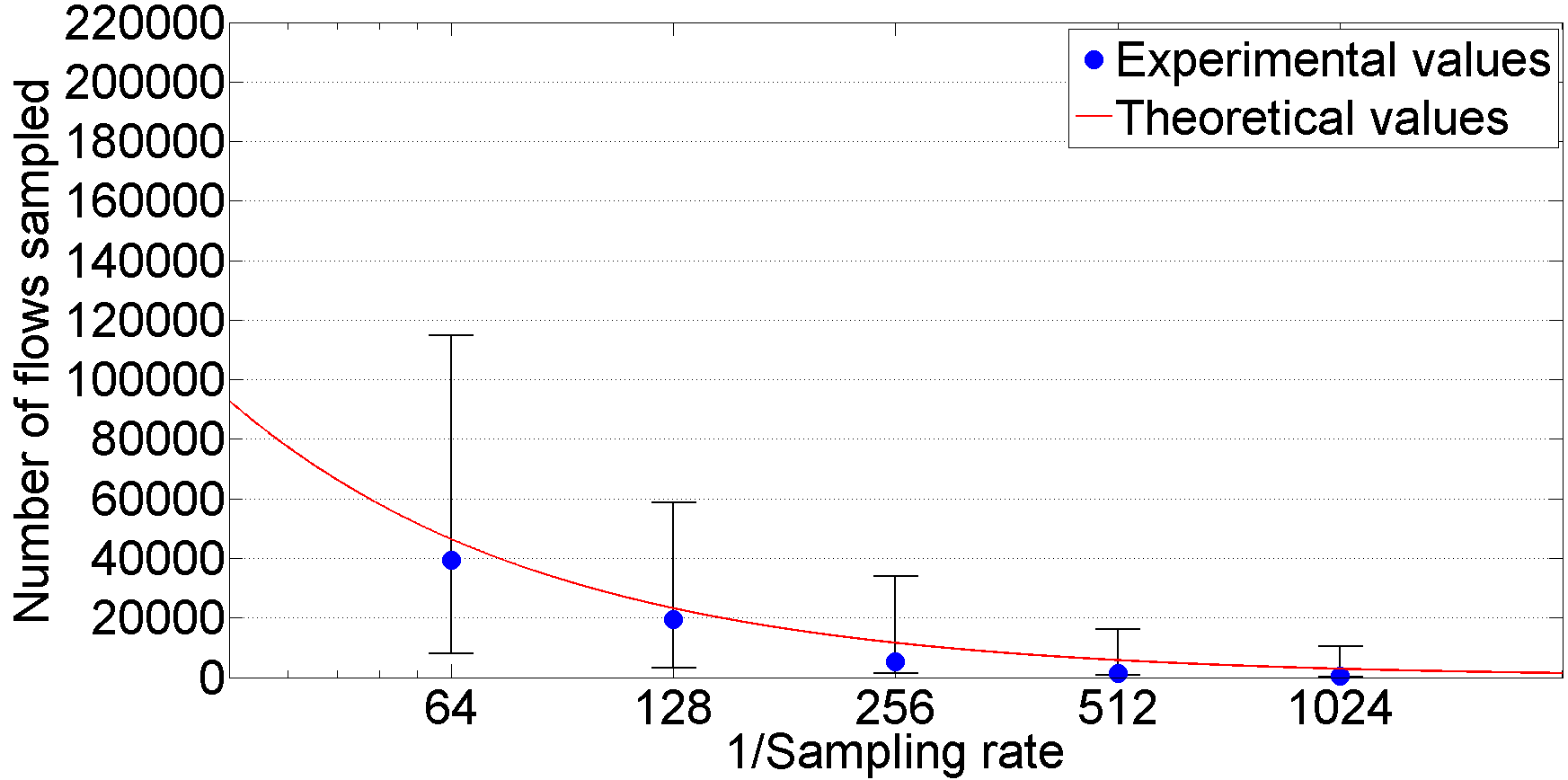}
  \caption{UNIVERSITY trace}
  \label{fig:sample-rate-port-pair:UPC}
\end{subfigure}
\caption{Evaluation of sampling rate for method based on pairs of ports}
\label{fig:sample-rate-port-pairs}
\end{figure}

To analyze the accuracy applying the sampling rate, we evaluate the number of flows sampled by our methods and compare it with the theoretical number of flows if we used a perfectly random selection function. We show in Fig. \ref{fig:sample-rate-IP-source} the results for the method based only on source IP suffixes. These plots display the median value of the number of flows sampled for the experiments conducted in relation to the sampling rate applied. The experimental values include bars which show the interval between the 5th and the 95th percentiles of the total 500 measurements obtained for each case. Likewise, in Fig. \ref{fig:sample-rate-pairs}, we show the same results for the case that considers pairs of source and destination IP suffixes. Given these results, we can see that the median values obtained are quite close to the theoretical values, i.e., in the average case these methods apply properly the sampling rate established. However, we can see there is a high variability between experiments. This means that, depending on the IP suffixes selected, we can over- or under-sample. Similar results are obtained for the method based on ports. In Fig. \ref{fig:sample-rate-port-source} (based on source ports) and Fig. \ref{fig:sample-rate-port-pairs} (based on pair of ports), we can observe that the median values are less closer to the theoretical ones and the variability among experiments is a bit greater.

\begin{figure}[!ht]
\centering
\begin{subfigure}{0.33\textwidth}
  \centering
  \includegraphics[width=\linewidth, height=2.8cm]{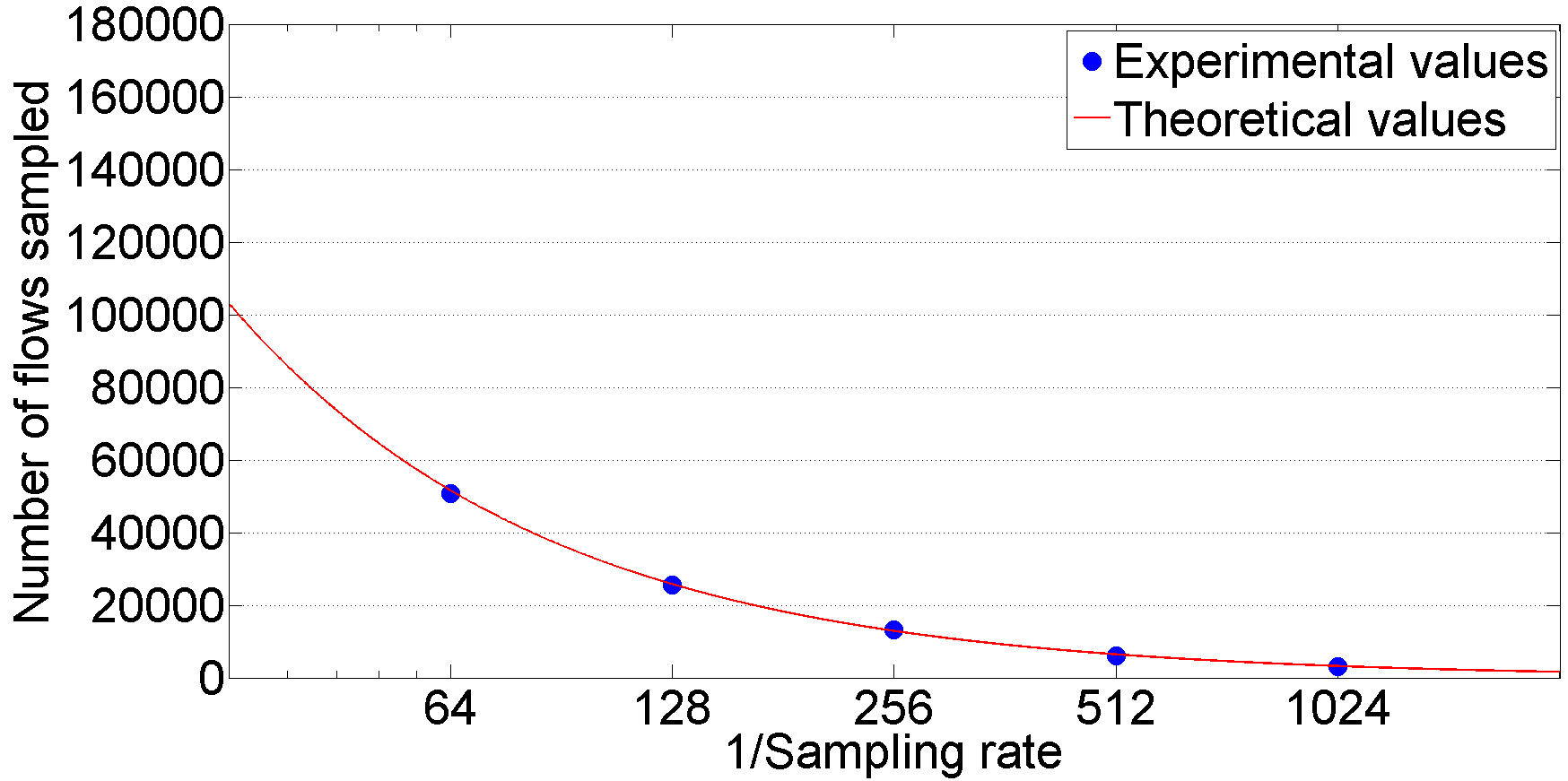}
  \caption{MAWI trace}
  \label{fig:sample-rate-hash-MAWI}
\end{subfigure}%
\begin{subfigure}{0.33\textwidth}
  \centering
  \includegraphics[width=\linewidth, height=2.8cm]{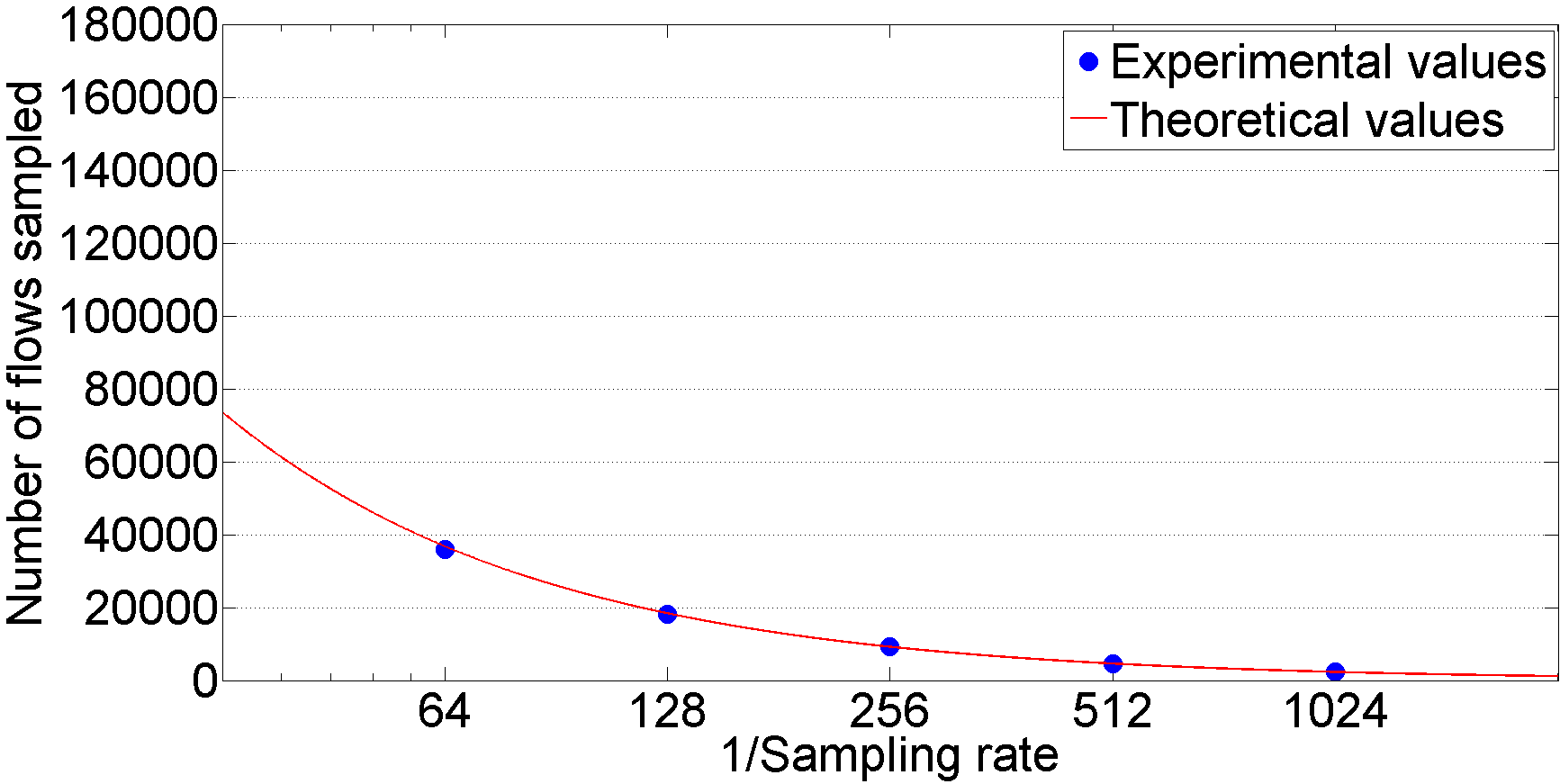}
  \caption{CAIDA trace}
  \label{fig:sample-rate-hash-CAIDA}
\end{subfigure}
\begin{subfigure}{0.33\textwidth}
  \centering
  \includegraphics[width=\linewidth, height=2.8cm]{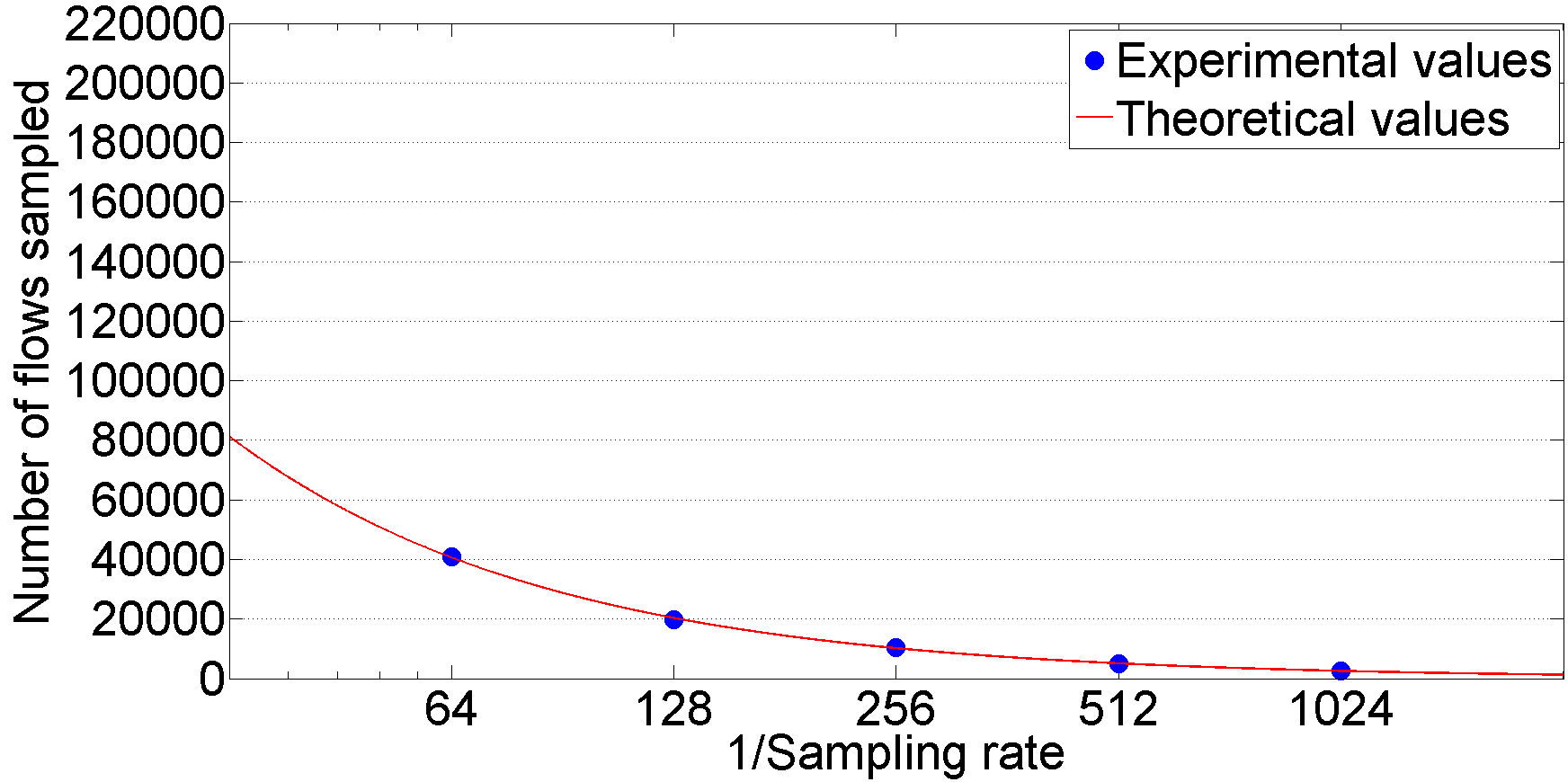}
  \caption{UNIVERSITY trace}
  \label{fig:sample-rate-hash-UPC}
\end{subfigure}
\caption{Evaluation of sampling rate for hash-based method}
\label{fig:sample-rate-hash}
\end{figure}

Next, we evaluate the hash-based sampling method making use of the load balancing algorithm for group tables included in Open vSwitch \cite{openvswitch}. The results in Fig. \ref{fig:sample-rate-hash} show that this method considerably outperforms the previous ones in terms of control of the sampling rate. Not only it samples a number of flows very close to the ideal one, but also it does not experience variability among experiments as it is based on a deterministic selection function.

In order to evaluate the randomness in the selection of our sampling methods, we compare our results with those obtained with a perfect implementation of flow sampling, with a completely random selection process. Thus, if our implementation is close to a perfect flow sampling implementation, the flow size distribution (FSD) should remain unchanged after applying the sampling, i.e., the distribution of the flow sizes (in number of packets) must be very similar for the original and the sampled data sets. Although we acknowledge that this property is not completely preserved for the methods based on IP suffixes and ports, we follow this approach to measure how random is the flow selection of this method and compare it with the hash-based method.

We quantify the randomness of the sampling method by calculating the difference between the FSDs of the original and the sampled traffic. For this purpose, we use the \textit{Weighted Mean Relative Difference} (WMRD) metric proposed in \cite{Duffield et al}. Thus, a small WMRD means that the flow selection is quite random. In Fig. \ref{fig:WMRD} we present boxplots with the results of our proposed methods. We can observe that these results are in line with the above results about the accuracy controlling the sampling rate. The method which shows better results is the hash-based one. Additionally, for the methods based on IP suffixes, we see that for the MAWI trace, the method based on pairs of IP suffixes achieves a more random flow subset. While for the CAIDA and UNIVERSITY traces, the method based on source IP suffixes behaves better.

\afterpage{%
\begin{figure}[!ht]
\vspace{-1cm}
\centering
\begin{subfigure}{.70\textwidth}
  \centering
  \includegraphics[width=\linewidth,height=3.5cm]{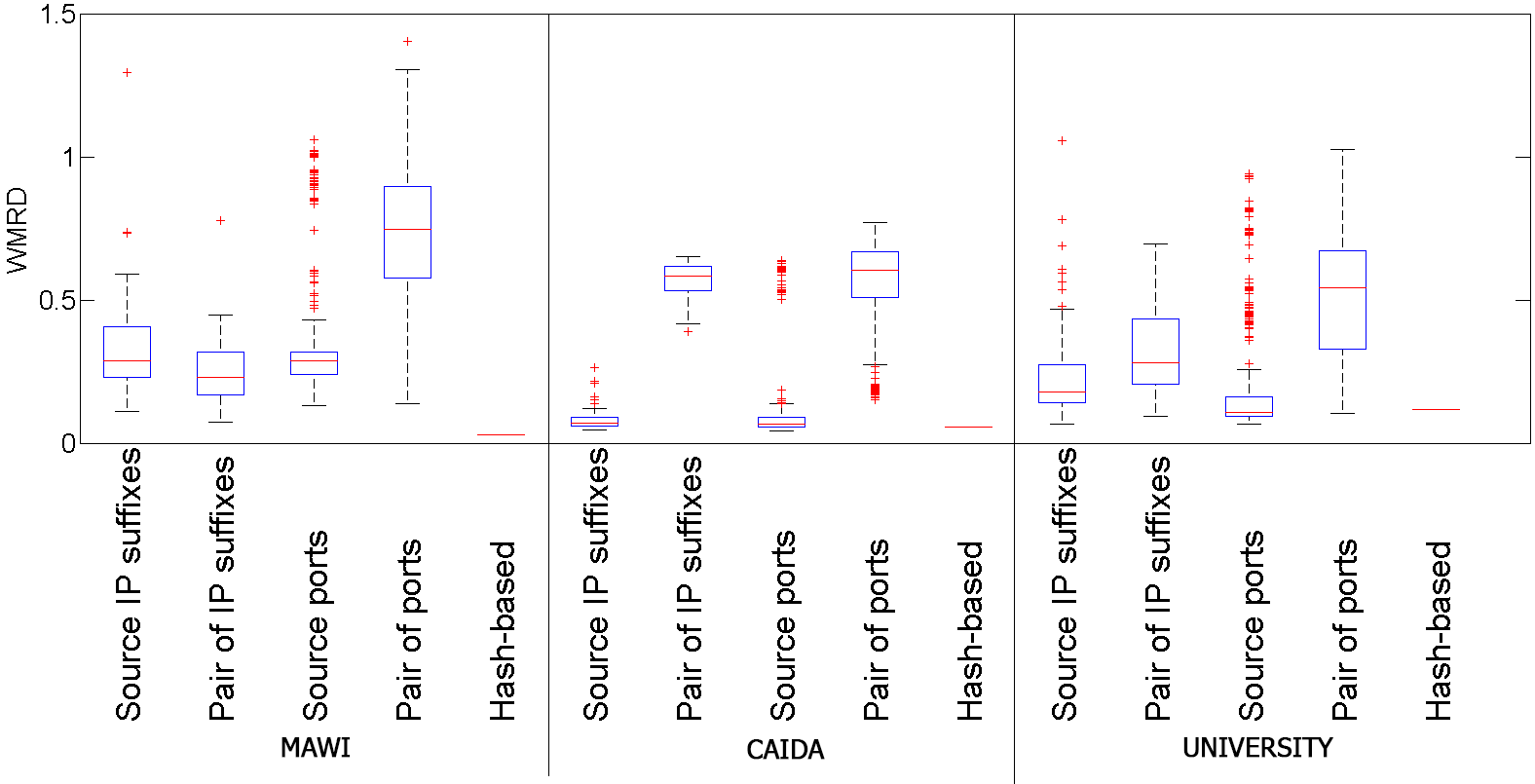}
  \caption{Sampling rate = 1/64}
   \label{fig:WMRD:64}
\end{subfigure}%
\hfil
\begin{subfigure}{.70\textwidth}
  \centering
  \includegraphics[width=\linewidth,height=3.5cm]{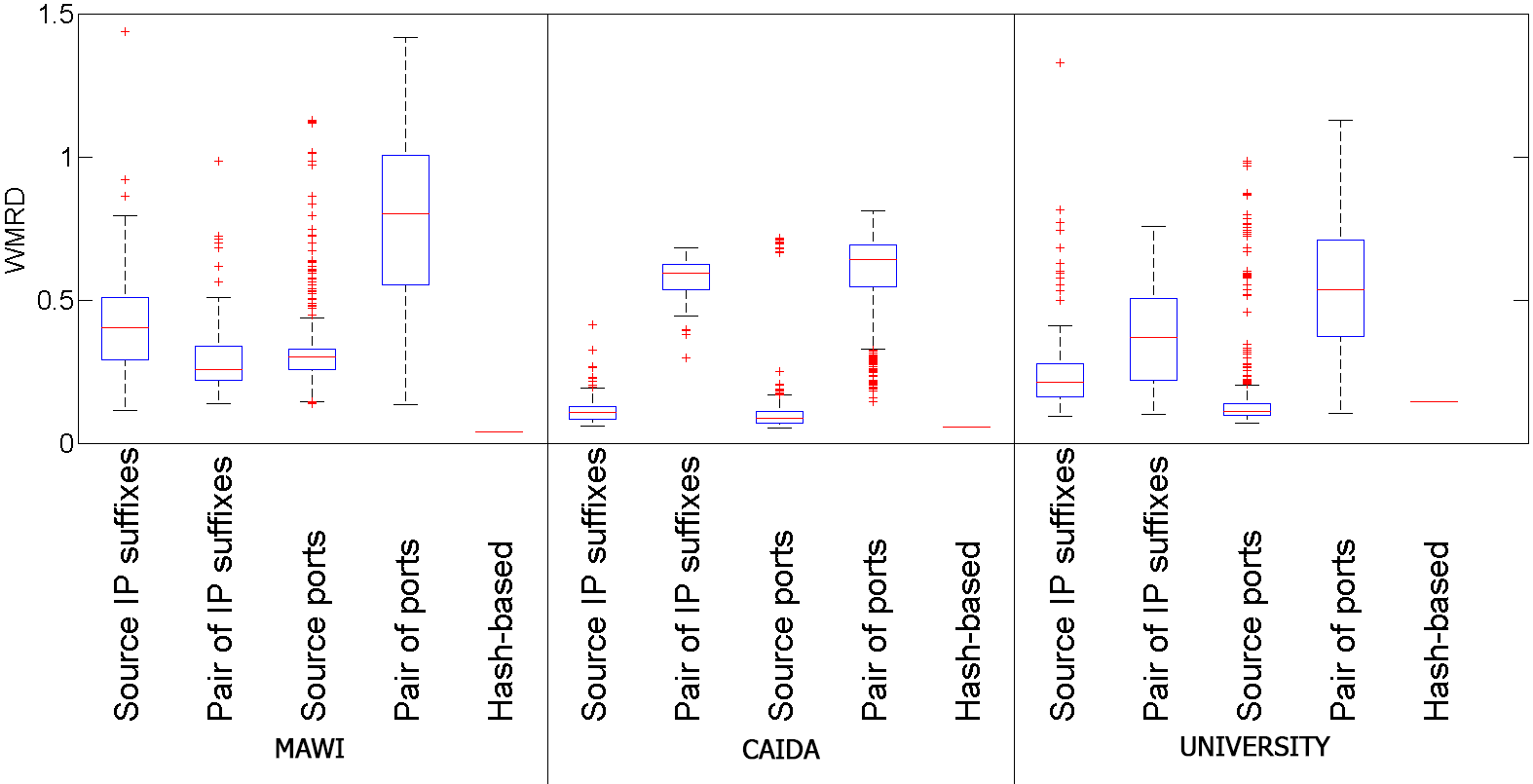}
  \caption{Sampling rate = 1/128}
   \label{fig:WMRD:128}
\end{subfigure}
\begin{subfigure}{.70\textwidth}
  \centering
  \includegraphics[width=\linewidth,height=3.5cm]{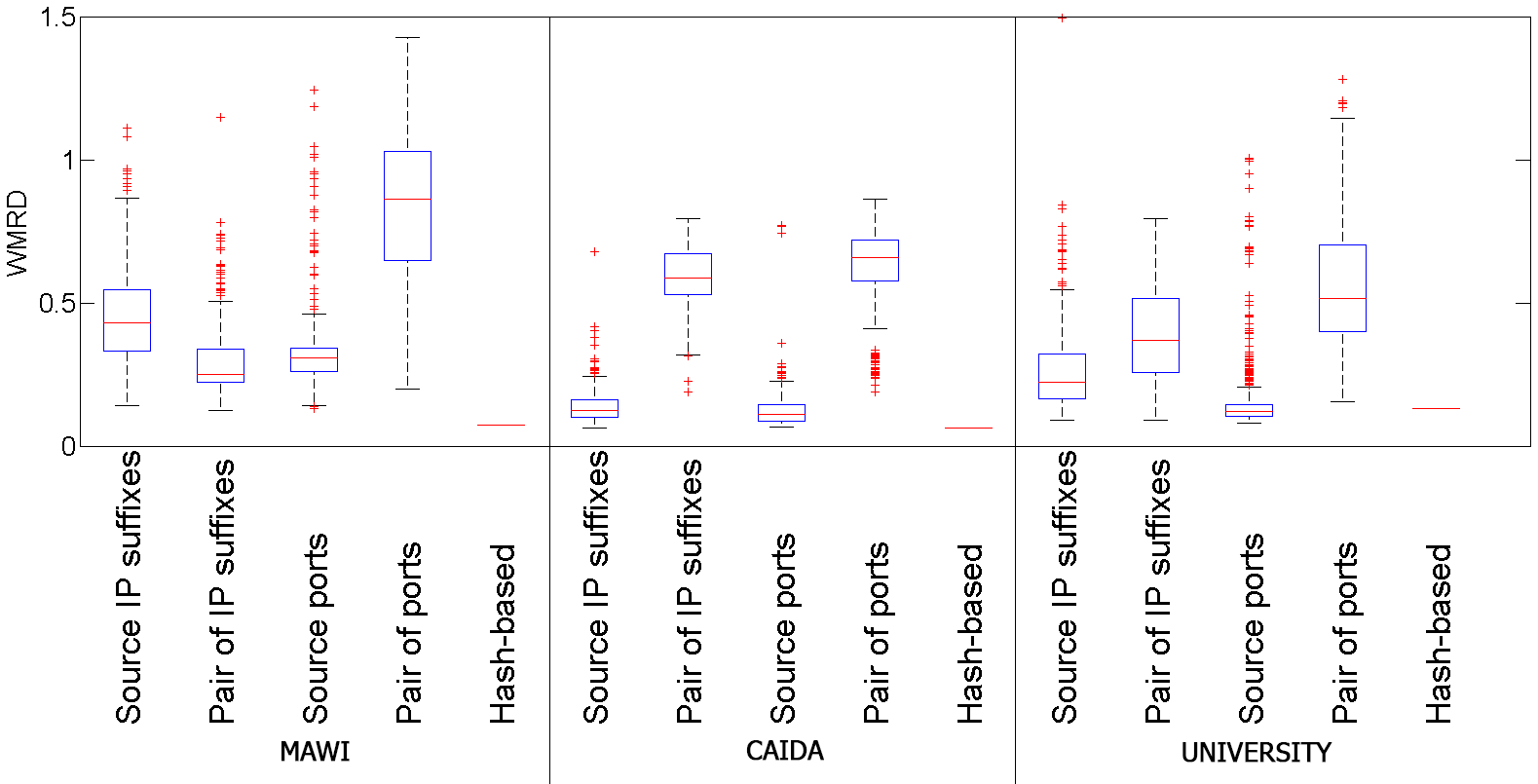}
  \caption{Sampling rate = 1/256}
  \label{fig:WMRD:256}
\end{subfigure}
\begin{subfigure}{.70\textwidth}
  \centering
  \includegraphics[width=\linewidth,height=3.5cm]{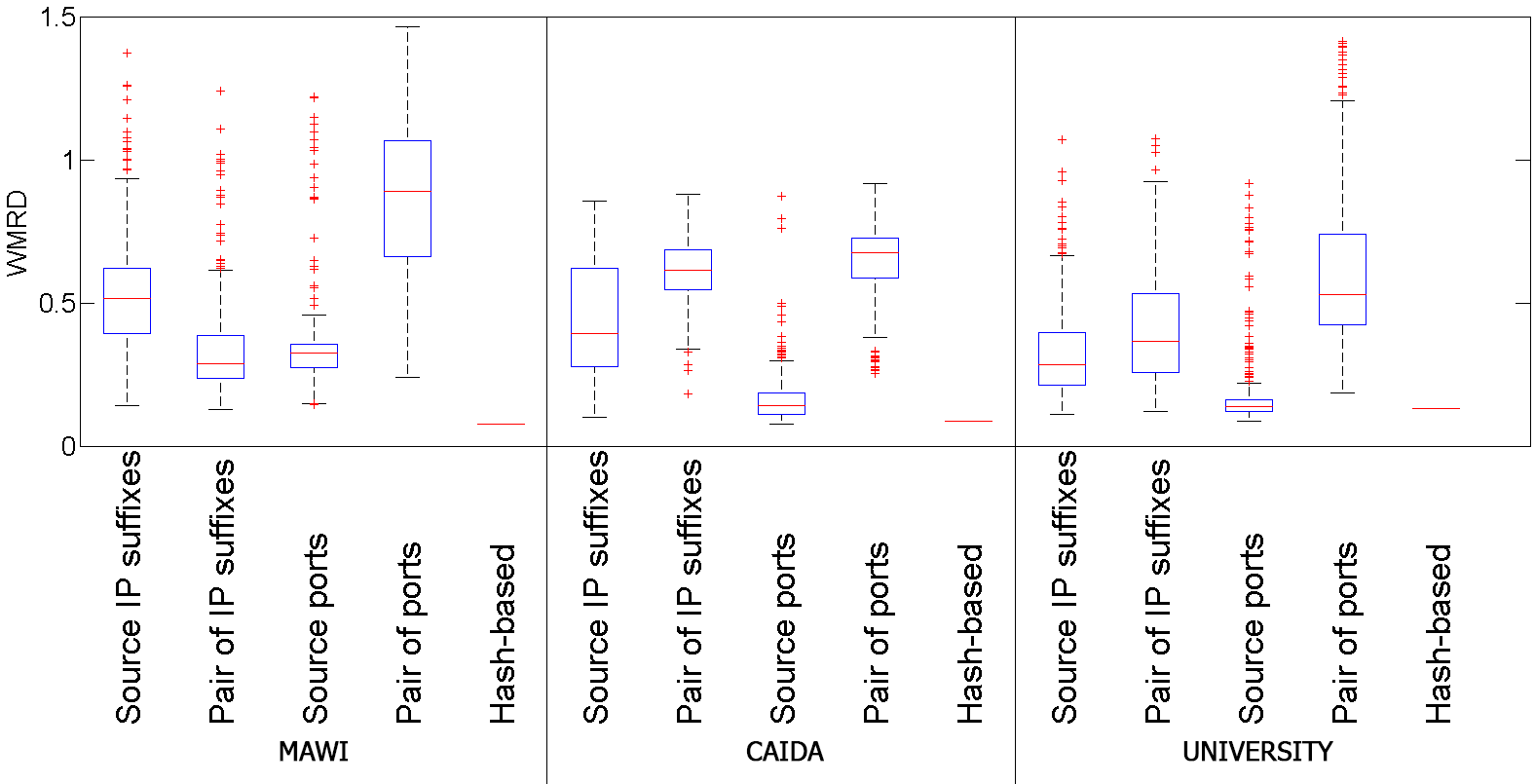}
  \caption{Sampling rate = 1/512}
  \label{fig:WMRD:512}
\end{subfigure}
\begin{subfigure}{.70\textwidth}
  \centering
  \includegraphics[width=\linewidth,height=3.5cm]{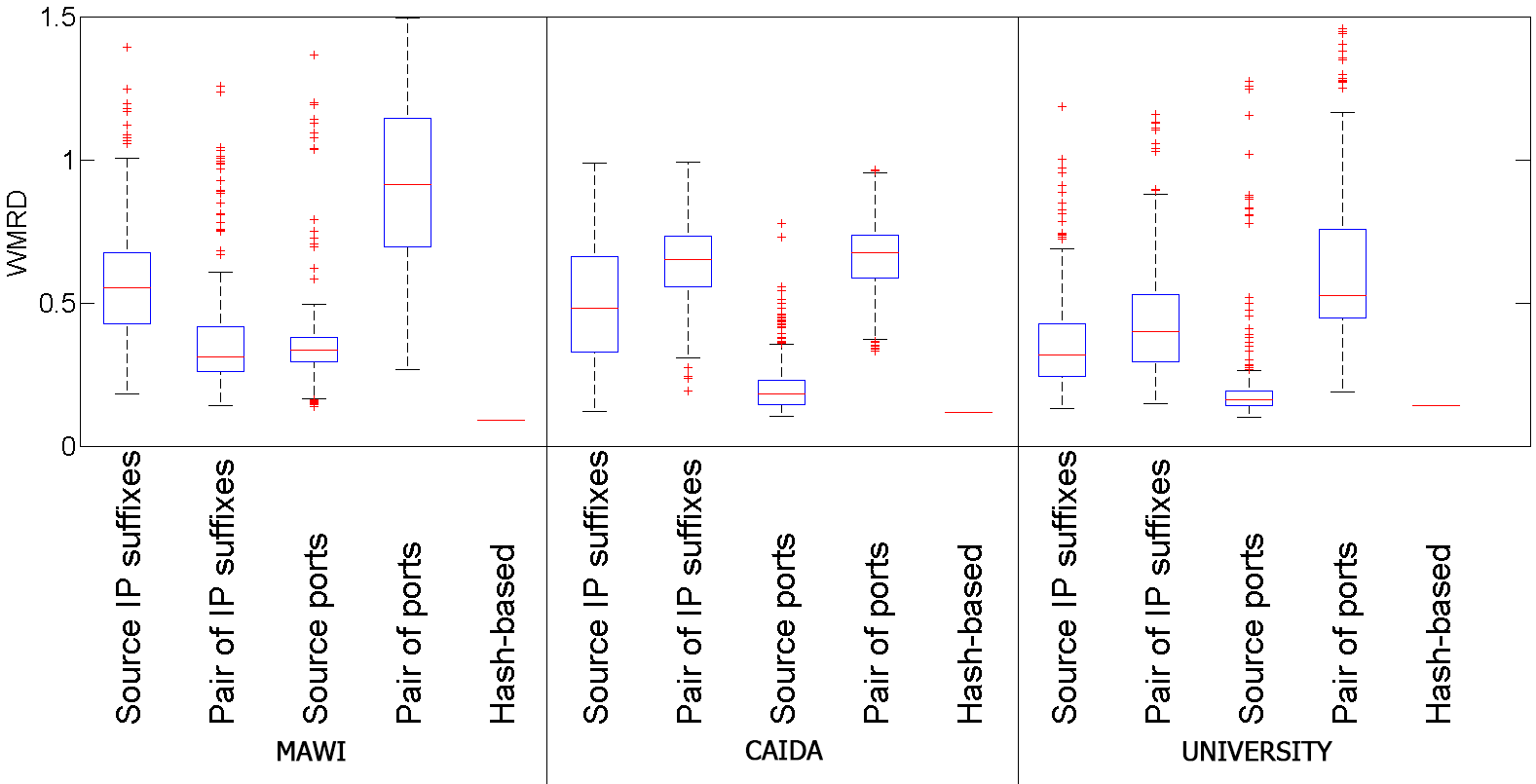}
  \caption{Sampling rate = 1/1024}
  \label{fig:WMRD:1024}
\end{subfigure}
\caption{Weighted Mean Relative Difference (WMRD) between FSDs}
\label{fig:WMRD}
\end{figure}
\clearpage
}

\begin{figure}[!ht]
\centering
\begin{subfigure}{0.49\textwidth}
  \centering
  \includegraphics[width=\linewidth, height=2.7cm]{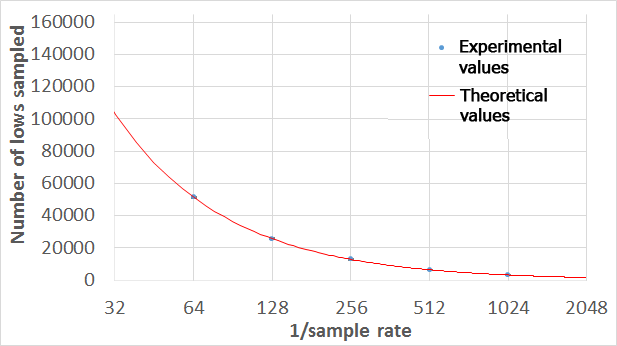}
  \caption{source IP suffix - MAWI trace}
  \label{fig:sample-rate-source-IP-MAWI-random}
\end{subfigure}%
\begin{subfigure}{0.49\textwidth}
  \centering
  \includegraphics[width=\linewidth, height=2.7cm]{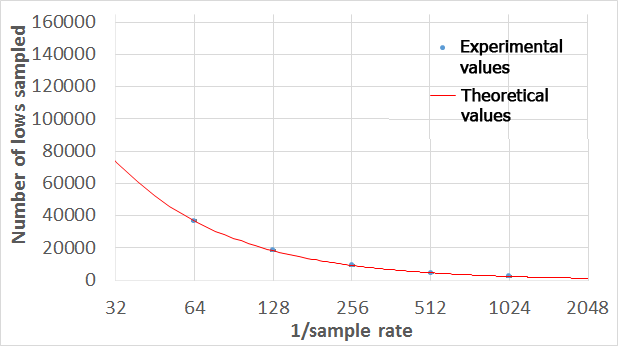}
  \caption{source IP suffix - CAIDA trace}
  \label{fig:sample-rate-source-IP-CAIDA-random}
\end{subfigure}
\begin{subfigure}{0.49\textwidth}
  \centering
  \includegraphics[width=\linewidth, height=2.7cm]{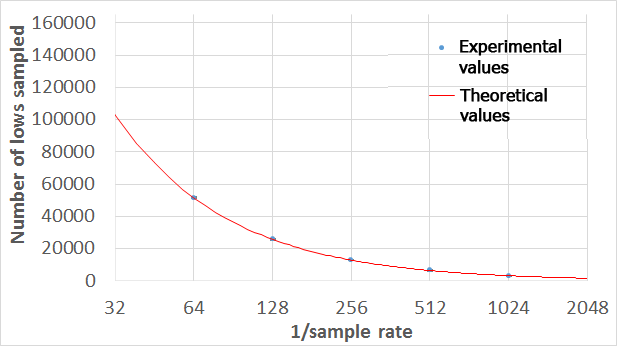}
  \caption{pair of IP suffixes - MAWI trace}
  \label{fig:sample-rate-pair-IP-MAWI-random}
\end{subfigure}
\begin{subfigure}{0.49\textwidth}
  \centering
  \includegraphics[width=\linewidth, height=2.7cm]{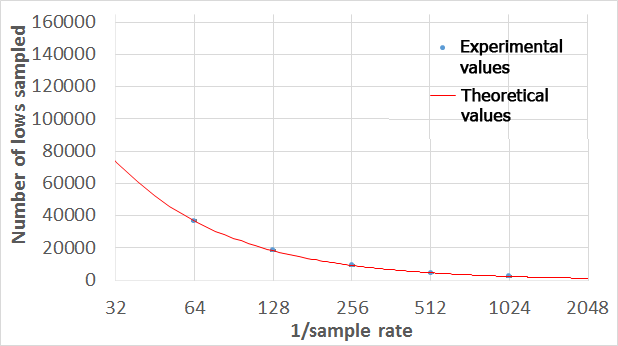}
  \caption{pair of IP suffixes - CAIDA trace}
  \label{fig:sample-rate-pair-IP-CAIDA-random}
\end{subfigure}
\caption{Evaluation of the method based on IP suffixes with randomized traces}
\label{fig:IP-suffixes-random}
\end{figure}

\begin{figure}[!ht]
\vspace{-1cm}
\centering
\begin{subfigure}{0.49\textwidth}
  \centering
  \includegraphics[width=\linewidth, height=2.7cm]{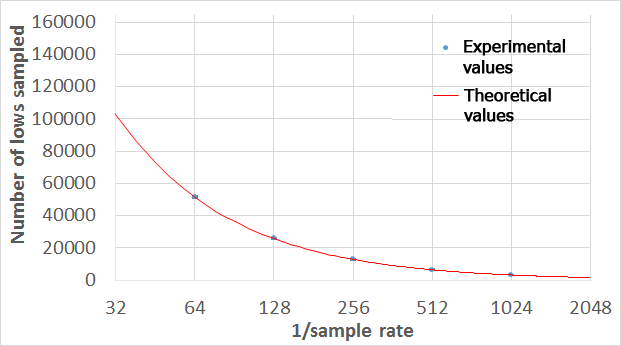}
  \caption{source ports - MAWI trace}
  \label{fig:sample-rate-source-port-MAWI-random}
\end{subfigure}%
\begin{subfigure}{0.49\textwidth}
  \centering
  \includegraphics[width=\linewidth, height=2.7cm]{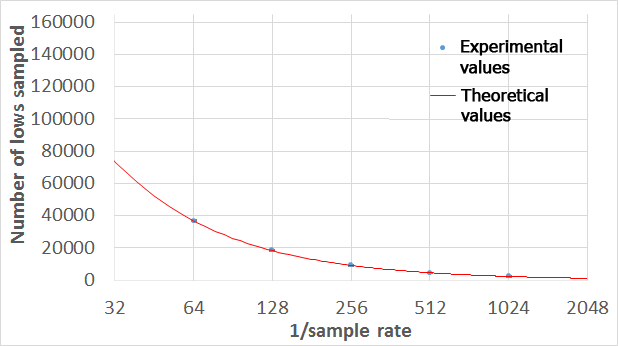}
  \caption{source ports - CAIDA trace}
  \label{fig:sample-rate-source-port-CAIDA-random}
\end{subfigure}
\begin{subfigure}{0.49\textwidth}
  \centering
  \includegraphics[width=\linewidth, height=2.7cm]{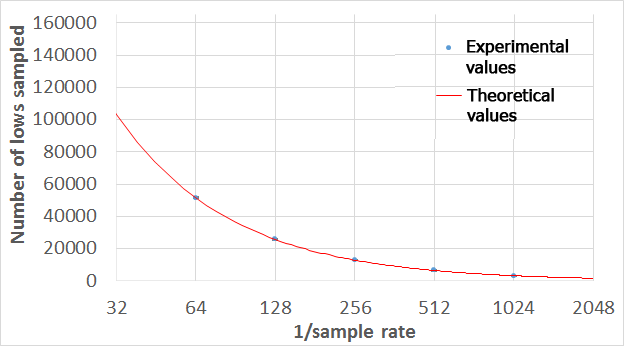}
  \caption{pair of ports - MAWI trace}
  \label{fig:sample-rate-pair-port-MAWI-random}
\end{subfigure}
\begin{subfigure}{0.49\textwidth}
  \centering
  \includegraphics[width=\linewidth, height=2.7cm]{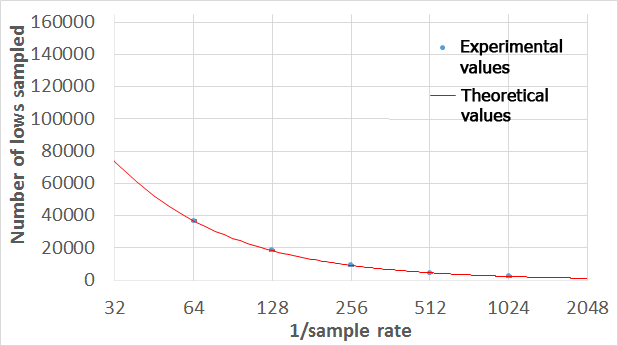}
  \caption{pair of ports - CAIDA trace}
  \label{fig:sample-rate-pair-port-CAIDA-random}
\end{subfigure}
\caption{Evaluation of the method based on ports with randomized traces}
\label{fig:port-suffixes-random}
\end{figure}

\begin{figure}[!ht]
\centering
\begin{subfigure}{.49\textwidth}
  \centering
  \includegraphics[width=\linewidth]{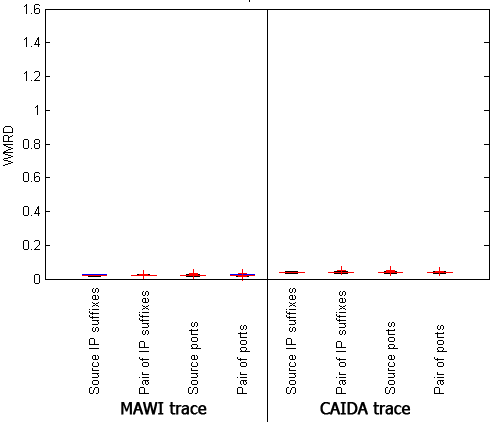}
  \caption{Sampling rate = 1/64}
   \label{fig:WMRD:64_random}
\end{subfigure}%
\begin{subfigure}{.49\textwidth}
  \centering
  \includegraphics[width=\linewidth]{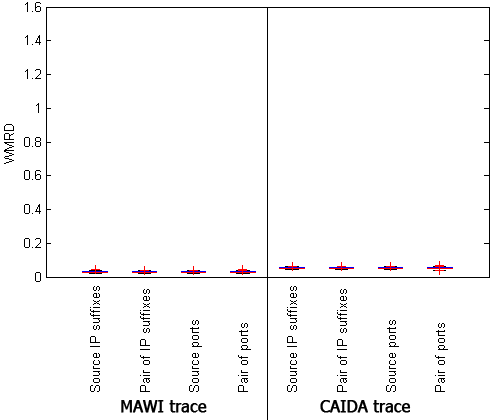}
  \caption{Sampling rate = 1/128}
   \label{fig:WMRD:128_random}
\end{subfigure}
\begin{subfigure}{.49\textwidth}
  \centering
  \includegraphics[width=\linewidth]{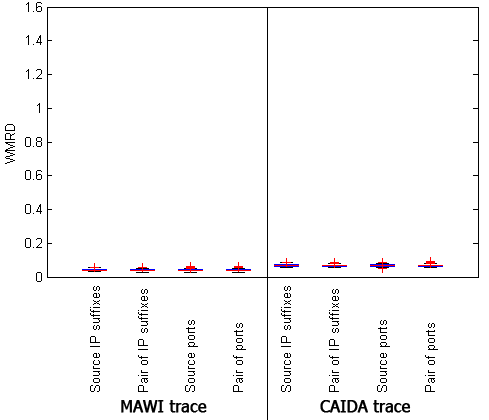}
  \caption{Sampling rate = 1/256}
  \label{fig:WMRD:256_random}
\end{subfigure}
\begin{subfigure}{.49\textwidth}
  \centering
  \includegraphics[width=\linewidth]{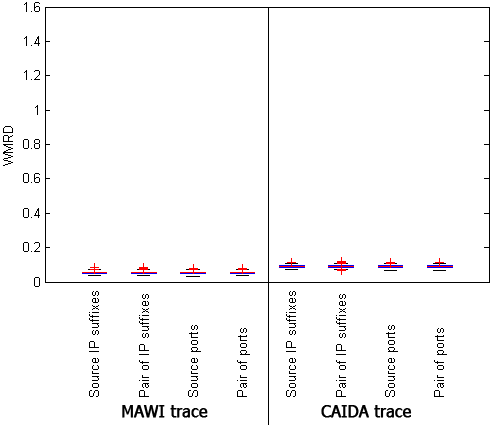}
  \caption{Sampling rate = 1/512}
  \label{fig:WMRD:512_random}
\end{subfigure}
\begin{subfigure}{.49\textwidth}
  \centering
  \includegraphics[width=\linewidth]{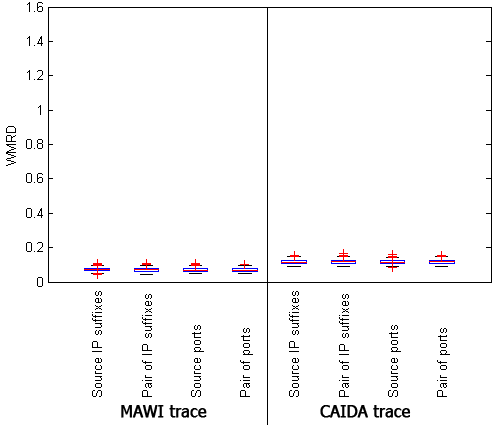}
  \caption{Sampling rate = 1/1024}
  \label{fig:WMRD:1024_random}
\end{subfigure}
\caption{Weighted Mean Relative Difference (WMRD) between FSDs with randomized traces}
\label{fig:WMRD_random}
\end{figure}

Lastly, we validate the implementation of the methods based on IP suffixes and ports.  We randomized the IPs and ports of the flows of the MAWI and CAIDA traces to have a homogeneous distribution and applied both methods. As for the control of the sampling rate, we present the results of the method based on IP suffixes (Fig. \ref{fig:IP-suffixes-random}), and based on ports (Fig. \ref{fig:port-suffixes-random}). Thus, we observe that for all the cases it achieved a number of flows very close to the theoretical values and a very small variability among experiments. In terms of randomness when selecting the flows, in Fig. \ref{fig:WMRD_random} we show the results for all the methods. We can observe that the WMRD is very low in all the cases if we compare it with the results obtained in Fig. \ref{fig:WMRD} with the original traces. These results also show that our methods based on IP suffixes and ports would work ideally in these conditions.

\newpage
\subsection{Evaluation of the overhead}

An inherent problem in OpenFlow is that, when we install flows reactively, packets belonging to the same flow are sent to the controller until a specific entry for them is installed in the switch. This is a common problem to any system that works at flow-level granularities. As a consequence, in our system we can receive in the controller more than one packet for each flow to be sampled. Specifically this occurs during the interval of time between the reception of the first packet of a flow in the switch, and the time when a specific entry for this flow is installed in the switch. This time interval is mainly the result of two factors: (I) the \textit{Round-Trip Time} (RTT) between the switch and the controller, and (II) the processing time of the controller to process the Packet In and execute the order to install a new entry. The RTT depends on some aspects like the distance between the switch and the controller or the capacity and utilization of the control link that connects them. The second factor depends on the processing power and the workload of the controller and, of course, its availability.

\begin{figure}[!ht]
\centering
\begin{subfigure}{0.49\textwidth}
  \centering
  \includegraphics[width=\linewidth, height=3cm]{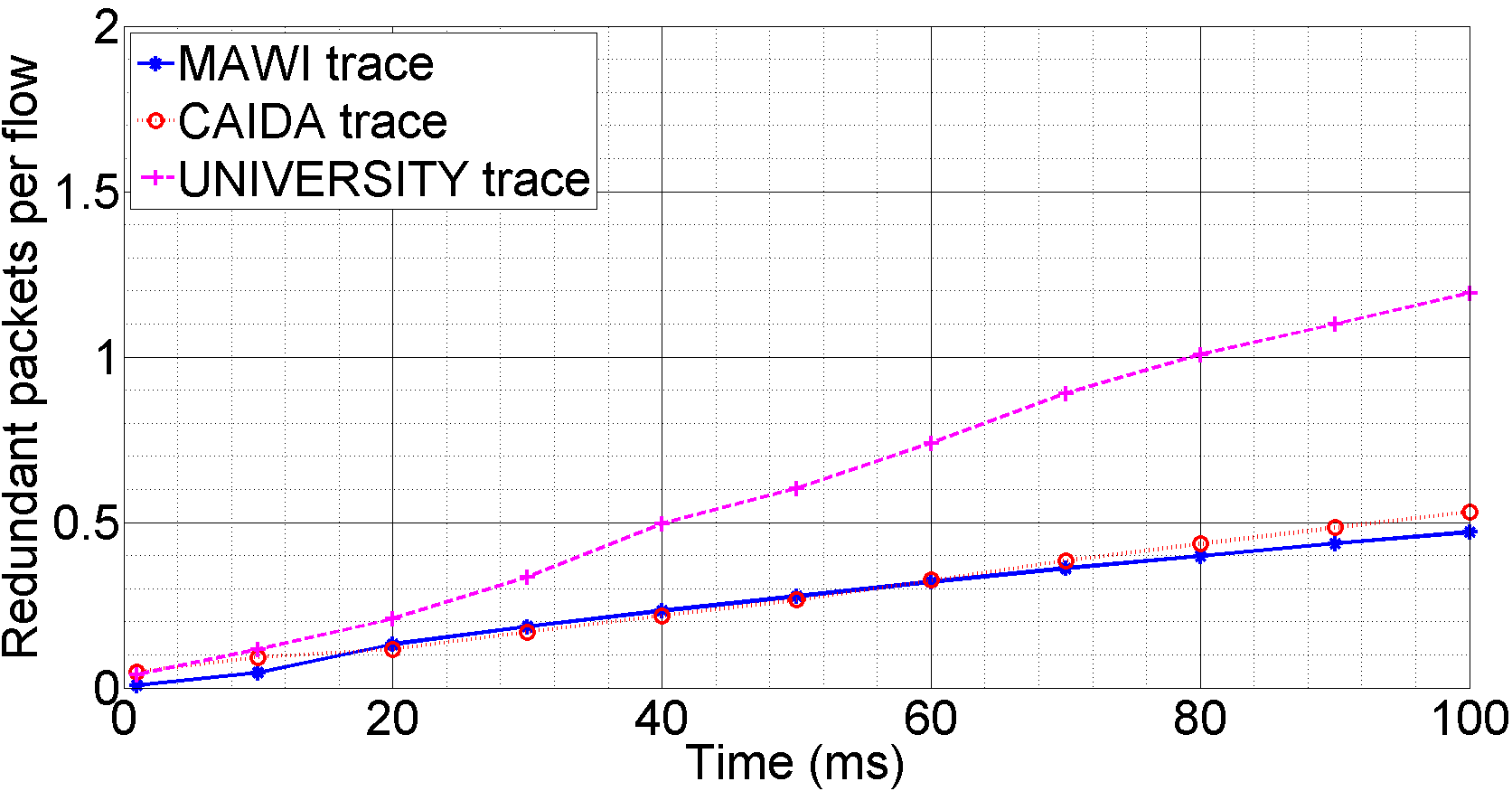}
  \caption{TCP traffic}
  \label{fig:redundant_packets_tcp}
\end{subfigure}%
\begin{subfigure}{0.49\textwidth}
  \centering
  \includegraphics[width=\linewidth, height=3cm]{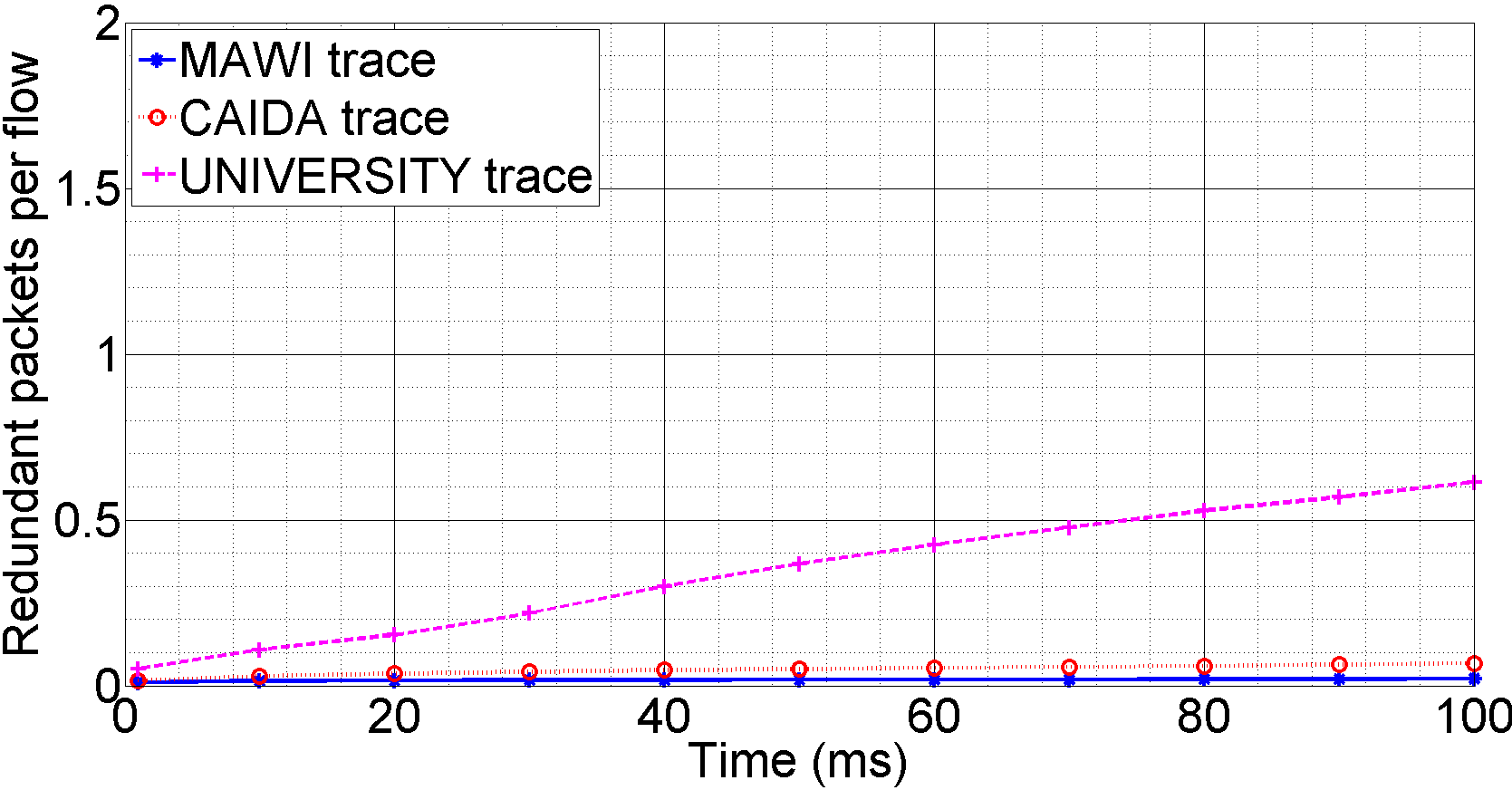}
  \caption{UDP traffic}
  \label{fig:redundant_packets_udp}
\end{subfigure}
\caption{Average number of redundant packets per flow.}
\label{fig:redundant_packets}
\vspace{-0.5cm}
\end{figure}

In order to estimate the amount of redundant packets of the same flow, we simulate an scenario where we consider a range from 1 ms to 100 ms for the elapsed time to install a new flow entry. As a reference, in \cite{onos} they observe a median value of 34.1 ms for the time interval to add a new flow with the ONOS controller in an emulated network with 206 software switches and 416 links. Thus, we simulate this range of time values for the three traces described in Table \ref{table:traces} and analyze the timestamps of the packets to calculate, for each flow, how many packets are within this interval and so would be sent to the controller. We analyze separately the overhead for TCP and UDP, as their results may differ due to their different traffic patterns. We show the results in Fig. \ref{fig:redundant_packets}. As we can see, the average number of redundant packets varies from less than 0.2 packets for delays below 20 ms, to approximately 1.2 packets per flow for an elapsed time of 100 ms for TCP traffic.

\begin{figure}[!ht]
\centering
\begin{subfigure}{0.49\textwidth}
  \centering
  \includegraphics[width=\linewidth, height=3cm]{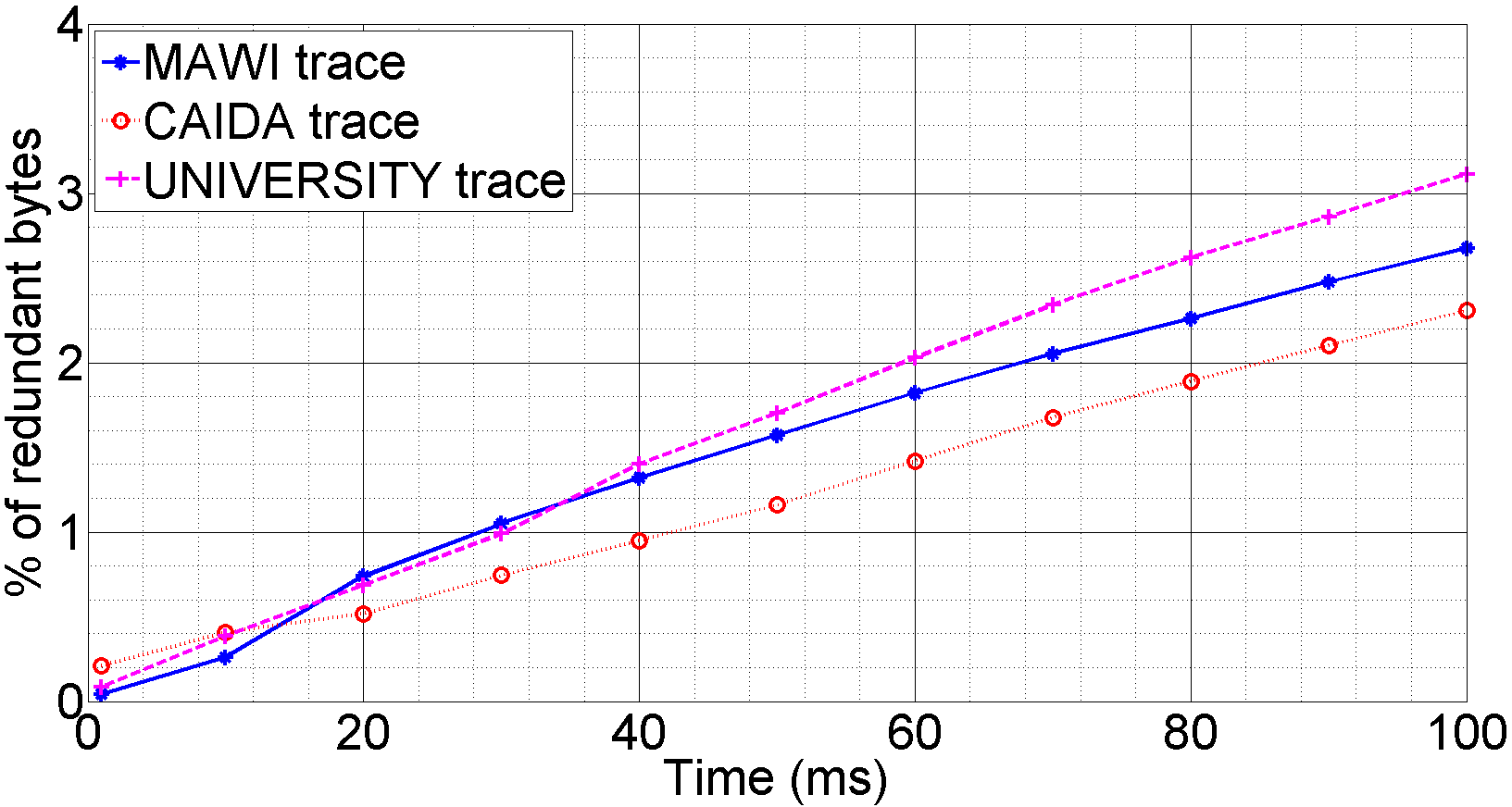}
  \caption{TCP traffic}
  \label{fig:redundant_bytes_tcp}
\end{subfigure}%
\begin{subfigure}{0.49\textwidth}
  \centering
  \includegraphics[width=\linewidth, height=3cm]{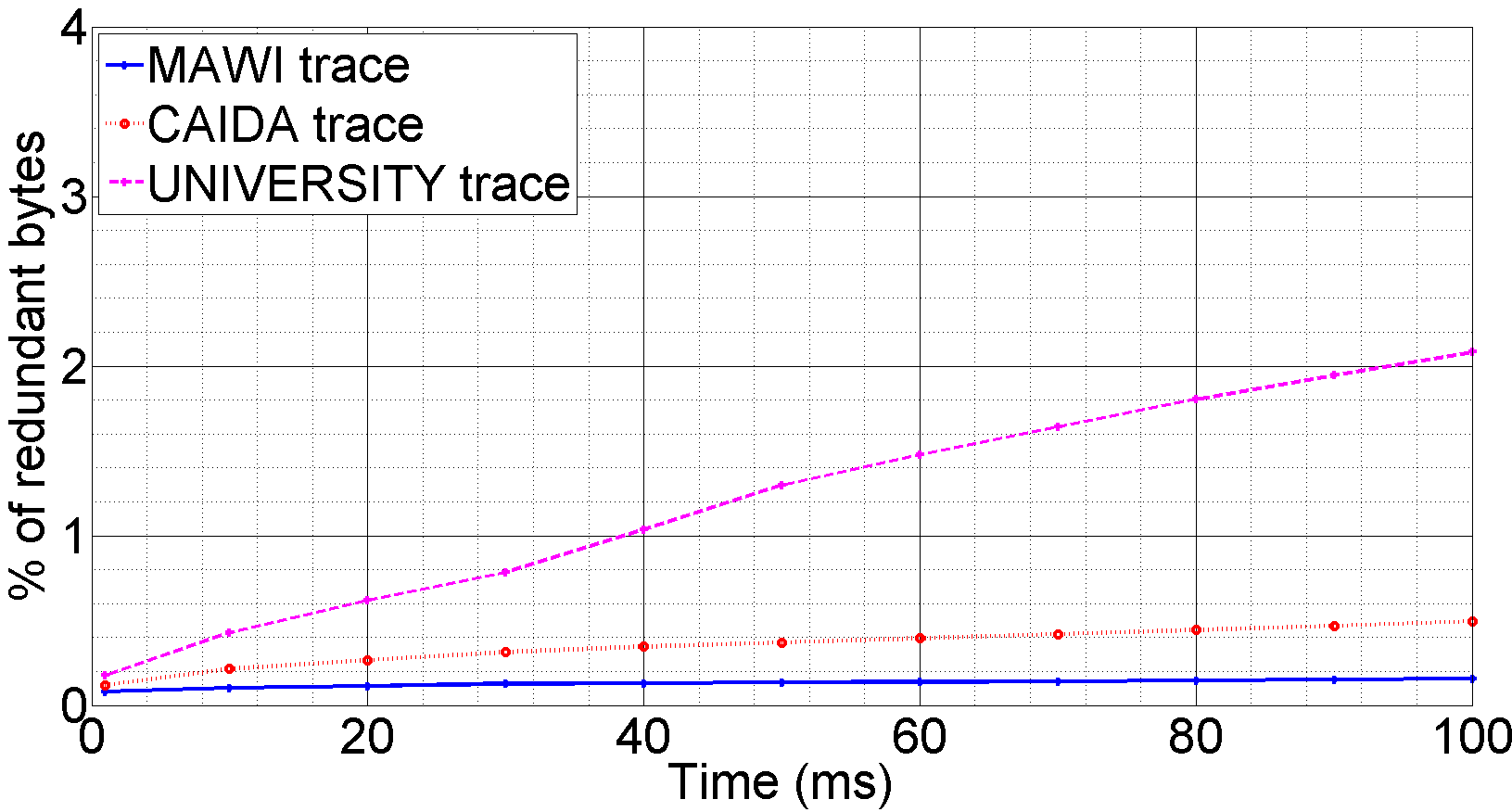}
  \caption{UDP traffic}
  \label{fig:redundant_bytes_udp}
\end{subfigure}
\caption{Percentage of redundant bytes.}
\label{fig:redundant_bytes}
\vspace{-0.5cm}
\end{figure}

Likewise, we show in Fig. \ref{fig:redundant_bytes} the results in terms of average percentage of redundant bytes sent to the controller. That way, the percentage of redundant bytes ranges from less than 0.8\% for elapsed times below 20 ms to 3.1\% in the worst case with an elapsed time of 100 ms and TCP traffic. These results show that the amount of redundant traffic sent to the controller is significantly smaller than if we implemented the trivial approach of forwarding all the traffic to the controller or a NetFlow probe and not installing in the switch specific entries to process subsequent packets and maintain per-flow statistics.

From these results we also infer that, for UDP traffic, the number of redundant packets and bytes per flow is significantly smaller than for TCP. Among other reasons, this is due to the fact that typically many UDP flows are single-packet (e. g. DNS request or responses). In the UNIVERSITY trace we could notice that there were more UDP flows with a larger number of packets, as it is reflected in Figs. \ref{fig:redundant_packets_udp} and \ref{fig:redundant_bytes_udp}.

We also remark that we should maintain records of the packets received in the controller to combine them with the statistics retrieved from the switch, which does not include the redundant packets in their correspondent packet and bytes counters.

As for the memory overhead in the switch, we implement sampling methods that provide mechanisms to control the number of entries installed. Thus, if we detect there is a large amount of flow entries in the switch, we can set adaptively the sampling rate according to the formula (\ref{eq:ip-suffixes}) for the method based on IP suffixes, the formulas (\ref{eq:pair-ports}) or (\ref{eq:single-port}) for the method based on ports, or by re-configuring the weights of the buckets for the hash-based method.

\section{Conclusions and future work}
We presented a monitoring solution which emulates the NetFlow/IPFIX operation with OpenFlow. In order to reduce the overhead in the controller and the number of entries required in the switch, we proposed three traffic sampling methods that can be implemented in current switches without requiring any modification to the OpenFlow specification. We implemented them in OpenDaylight and evaluated their accuracy in a testbed with real traffic. As future work, we plan to implement smarter algorithms to adaptively select the timeouts in order to retrieve the statistics more accurately and also a packet sampling method, although we find it more challenging.

\section*{Acknowledgement}
This work was done in collaboration with Talaia Networks under the framework of the H2020 SME Instrument Phase 2 project "SDN-Polygraph" (ref. 726763). This work was also supported by the Spanish Ministry of Economy and Competitiveness and EU FEDER under grant TEC2014-59583-C2-2-R (SUNSET project) and by the Catalan Government (ref. 2014SGR-1427).

\newpage

\newpage

%
%
%

\end{document}